\documentclass{bmcart}
\usepackage[utf8]{inputenc} 
\usepackage{indentfirst}
\usepackage{natbib}
\usepackage{ boldline, rotating}
\usepackage{array}
\usepackage[export]{adjustbox}
\usepackage{makecell,multirow}
\usepackage{amssymb}
\usepackage{mathtools}
\usepackage{hyperref}
\usepackage{bigfoot}
\usepackage{mwe}
\usepackage{amsmath}

\usepackage{afterpage}

\usepackage{graphicx}
\usepackage{lipsum} 
\usepackage{wrapfig} 
\newlength{\tempdima}
\newcommand{\rowname}[1]
{\rotatebox{90}{\makebox[\tempdima][c]{\textbf{#1}}}}

\startlocaldefs
\endlocaldefs

\begin{document}

\begin{frontmatter}

\begin{fmbox}
\dochead{Research}


\title{Investigating the contribution of author- and publication-specific features to scholars' h-index prediction}


\author[
   addressref={aff1},                   
   corref={aff1},                       
   email={fakhri.momeni@t-online.de}   
]{\inits{}\fnm{Fakhri} \snm{Momeni}}
\author[
   addressref={aff1},
   email={philipp.mayr@gesis.org}
]{\inits{}\fnm{Philipp} \snm{Mayr}}
\author[
   addressref={aff1,aff2},
   email={stefan.dietze@hhu.de}
]{\inits{}\fnm{Stefan} \snm{Dietze}}


\address[id=aff1]{
  \orgname{GESIS -- Leibniz Institute for the Social Sciences}, 
  \street{Unter Sachsenhausen 6-8},                     %
  \postcode{50667}                                
  \city{Cologne},                              
  \cny{Germany},                                    
}
\address[id=aff2]{%
  \orgname{Heinrich-Heine-University },
  \street{Universit\"{a}tsstr. 1},
  \postcode{40225}
  \city{D\"{u}sseldorf},
  \cny{Germany}
}


\begin{artnotes}
\end{artnotes}

\end{fmbox}


\begin{abstractbox}

\begin{abstract} 
Evaluation of researchers' output is vital for hiring committees and funding bodies, and it is usually measured via their scientific productivity, citations, or a combined metric such as the h-index. Assessing young researchers is more critical because it takes a while to get citations and increment of h-index. Hence, predicting the h-index can help to discover the researchers' scientific impact. In addition, identifying the influential factors to predict the scientific impact is helpful for researchers and their organizations seeking solutions to improve it. This study investigates the effect of the author, paper/venue-specific features on the future h-index. For this purpose, we used a machine learning approach to predict the h-index and feature analysis techniques to advance the understanding of feature impact. Utilizing the bibliometric data in Scopus, we defined and extracted two main groups of features. The first relates to prior scientific impact, and we name it 'prior impact-based features' and includes the number of publications, received citations, and h-index. The second group is 'non-prior impact-based features' and contains the features related to author, co-authorship, paper, and venue characteristics. We explored their importance in predicting researchers' h-index in three career phases. Also, we examined the temporal dimension of predicting performance for different feature categories to find out which features
are more reliable for long- and short-term prediction. We referred to the gender of the authors to examine the role of this author's characteristics in the prediction task. Our findings showed that gender has a very slight effect in predicting the h-index. Although the results demonstrate better performance for the models containing prior impact-based features for all researchers' groups in the near future, we found that non-prior impact-based features are more robust predictors for younger scholars in the long term. Also, prior impact-based features lose their power to predict more than other features in the long term.
\end{abstract}




\end{abstractbox}
%

\end{frontmatter}




\section{Introduction}\label{secIn}
Predicting scientific impact helps to anticipate the career trajectories of researchers and reveal mechanisms of the scientific process that influence future impact, which has always been a concern of individual researchers, universities, recruitment committees, and funding agencies. Also, it can reveal factors influencing the future outcome and propose path-ways to young researchers on how to improve future impact and their organizations for more support.

Scientific productivity and received citations are the basis for many evaluation metrics (e.g., h-index \cite{hirsch2005index},  g-index \cite{egghe2006improvement}, $h_{s}$-index   \cite{kaur2013universality}). The h-index is the most common metric which evaluates the scholars' scientific impact since it measures researchers' productivity and citation impact and has a leading role in hiring and funding decisions. Therefore, predicting this metric is crucial for these purposes. The shorter publication record, received citations, and h-index (prior impact-based features) simplify the h-index prediction task because these features reflect the scholar's impact.
Since more senior scholars have a distinguished research profile, predicting their h-index is easier. Assessing the future impact is more pivotal for young scholars than seniors because prior impact-based features are less available for junior researchers as they have a shorter data history.
The prediction task will be more complicated for rising stars (who have a lower research profile at the beginning of their career compared to other authors in the same career stage but may become prominent contributors in the future \cite{daud2013finding}), and we need non-prior impact-based features to evaluate their impact in the long term.
Although previous studies demonstrated high accuracy by employing prior impact-based features \cite{ayaz2018predicting,weihs2017learning,wu2019predicting}, they displayed a substantial decline in the performance of predicting the h-index in the distant future. We hypothesise that publication/citation-based features may be efficient short-term predictors, but other feature categories may be more efficient in predicting long-term impact.

To address these limitations and improve the accuracy of h-index prediction, this study takes a comprehensive approach by investigating a wide array of features and feature sets. We consider traditional publication/citation-based features and explore other feature categories that may play a role in predicting long-term impact. Our primary objective is to gain a deeper understanding of feature contributions to the h-index prediction task for researchers at different career stages. Our investigation involves analyzing various features and feature sets in the context of h-index prediction. Drawing from prior research associating specific features with productivity and received citations, we examine how these attributes contribute to researchers' future h-index.  To accomplish this, we leverage a machine learning approach to predict the h-index for the upcoming ten years and conduct an extensive feature analysis. To assess the temporal stability of our predictions, we implement our method on three distinct groups of authors: junior, middle-level, and senior researchers. By comparing the accuracy of different feature combinations within each group, we gain insights into the efficacy of the predictive models over time.

In summary, our study makes three significant contributions to the field:
\begin{enumerate}
    \item \textbf{Feature impact analysis:} We advance the understanding of the impact of different feature categories on various h-index prediction tasks for researchers in different career phases and examine the reliability of these predictions.
    \item \textbf{Temporal dimension of feature performance:} We investigate the temporal dimension of predictors to advance the understanding of feature performance depending on the time window considered for the future prediction, i.e., to understand which features/categories perform better for long- and short-term prediction regarding their seniority.
    \item \textbf{Novel features:} We introduce and investigate the effect of non-prior impact-based features, namely gender and academic mobility, on the prediction task to reveal the influential factors on the scientific impact (prior impact-based features that implicitly or explicitly encode citation counts simplify the h-index prediction task dramatically by providing the model with data that directly influences the target metric (h-index)).  
\end{enumerate}

\section{Related Work}\label{secRe}
 To identify the future scientific impact, several studies focus on predicting the citations count for a specific paper \cite{bai2019predicting, abrishami2019predicting, jiang2021hints, ruan2020predicting, kossmeier2019predicting}, others tried to predict the impact at the author level with the h-index \cite{ayaz2018predicting, weihs2017learning, wu2019predicting, nikolentzos2021can}. 
Among all models and methods presented in these studies to predict the h-index, those that took the number of prior publications, received citations, or the current h-index (prior impact-based features) into consideration achieved the highest performance.
Although prior impact-based features are the strongest predictors of future impact, sometimes we need to predict it using the other author, paper, and venue characteristics. 
\subsection{Features Used for the Prediction Tasks}
Many studies employed various properties of papers, venues, authors, and their coauthors to predict the scientific impact. Abrishami and Aliakbary \cite{abrishami2019predicting} and Bai et al. \cite{bai2019predicting} use time series methods and early citations count to predict the number of citations in the long term. Jiang et al. \cite{jiang2021hints} presented a citation time series approach to predict the citations for newly published papers. They used the paper's topic (via keyword), author reputation, venue prestige, and temporal cues (e.g., increasing network centrality over time) to detect citation signals and convert them into signals for citation time series
generation. Nie et al. \cite{nie2019academic} utilized some features and categorized them into the author (regarding citations and publication), venue, social (coauthor), and temporal (average citation increment of the author and coauthors within two years) features and examined their importance in predicting academic rising star.   
 Ayaz et al. \cite{ayaz2018predicting} and Weihs and Etzioni \cite{weihs2017learning} used the number of current publications, citations, or h-index with other features to predict the future h-index and both presented models with ${R^2=0.93}$. Wu et al. \cite{wu2019predicting} included related indicators to these features, such as changes in citations and h-index over the last two years to the predictors' list and demonstrated a model with a higher precision ${R^2=0.97}$. Further studies focused on other feature types rather than prior impact-based features to identify the influential factors on the scientific impact of researchers. For example, McCarty et al. \cite{mccarty2013predicting} investigated the relationship between some characteristics of the coauthor network and the h-index. Their results showed the significance of coauthors' productivity via collaborating with many authors and their impact on predicting the h-index. Nikolentzos et al. \cite{nikolentzos2021can} extracted two types of features, papers' textual content and graph features (related to collaboration patterns), and found that graph features alone are more robust predictors. Dong et al. \cite{dong2016can} studied the contribution of a publication to the author's h-index and found that topical authority and publication venues are the most predictive features in the absence of citation-related features of prior publications. Otherwise, they reported citation count as the most decisive factor in predicting the future h-index. Jiang et al. \cite{jiang2021hints} found that certain features, such as the author's reputation, are more predictive than others. Therefore, they applied trainable weights to preserve the unequal contribution of different kinds of features.
 Ayaz et al. \cite{ayaz2018predicting} reported the career age, number of high-quality papers, and number of publications in distinct journals as the most compelling feature in predicting the h-index after prior impact-based features. They observed a lower performance for younger researchers and concluded the investigated features are insufficient to predict their h-index and a need to evaluate future features for better prediction. 

 Wu et al. \cite{wu2019predicting} investigated the stability of predictive models for long-term prediction (ten future years) and compared  their method with state-of-the-art \cite{ayaz2018predicting,weihs2017learning,dong2016can}. They used time series features (the history of the h-index)  and more impact-based features in their analyses, which are less valuable to predict the future impact of young researchers. They found better performance among all mentioned works. However, they included only the authors with an h-index higher than four and junior researchers whose predicting their scientific impact is more challenging have been excluded from their study. 
 
 We tackle these issues by investigating novel author- and paper-specific features for the prediction task and verifying their contribution to the h-index prediction for researchers with varying scientific experiences.

\subsection{Influential Factors on Scientific Impact}\label{subsecI}
In the following, we categorize the features affecting the scientific impacts into three groups: demographic, paper/venue, and coauthor-based factors, and report the previous related studies. 
\subsubsection{Demographic Factors}
\textbf{Academic mobility:} In contemporary science, collaboration plays a significant role, and international academic mobility affects the collaboration networks, which furthers knowledge transmission among countries and scholars. Therefore, many studies have focused on investigating its impact on science and scientists. Our recent study \cite{momeni2022many} revealed the positive impact of international mobility on the number of publications and received citations. However, mobile researchers do not necessarily perform better than those without mobility experience. Singh \cite{singh2018comparing} found that differences in research outputs between returnee Ph.D. holders and those trained in their home country are field-specific and depend on their seniority. Netz et al. \cite{netz2020effects} reviewed the studies that investigated the effect of mobility on some scientific outcomes and found that most studies suggest a positive effect on mobility. But they reported some studies that demonstrated a negative effect on productivity and citation impact and proposed a positive impact of mobility only under specific circumstances. Liu et al. \cite{liu2021academic} found that international collaboration before mobility has an essential role in high performance after mobility. The reputation of institutions is another influential factor they discovered in their study.

\textbf{Gender:} Gender differences in science and scientific impact have been the subject of many studies in various fields. A new study on the Breast Surgery Fellowship Faculty \cite{radford2022h} found no noticeable gender difference between assistant professors but a higher h-index for men professors than women. \cite{carter2017gender} studied the gender gap in social sciences and found the difference in all career phases, especially in full professor positions. In contrast, the study's results by Lopez et al. \cite{lopez2014gender} demonstrated a higher h-index for men among academic ophthalmologists. Still, controlling the range of publications, they found the same or more impact for women in the later career phases. The results of the study by Kelly et al. \cite{kelly2006h} indicated that although the h-index of men is higher than women for
ecologists and evolutionary biologists, there is no gender difference in the h-index once we control for publication rate. However, other studies \cite{leydesdorff2019relative,smirnova2023comprehensive} examined the relationship between received citations and funding available from Web of Science data and found a weak correlation between them. 

\textbf{Income level:} In many countries, governments are the primary source of financial support for scientific progress. Gantman \cite{gantman2012economic} demonstrated the positive effect of economic development on scientific productivity in all scientific fields. 
Confraria et al. \cite{confraria2017determinants} displayed a U-shape relationship between Gross Domestic Product (GDP) per capita and received citations and found the citation impact correlates positively with the nation’s wealth after a certain GDP per capita level. However, their results showed that international collaboration is crucial for higher citation impact among all countries. 

\subsubsection{Paper and Venue Factors}\label{subsubsecR2}
\textbf{Scientific field}: The average scholars' h-index of researchers differs among fields because productivity and the rate of citing vary from one to another \cite{malesios2014comparison,lillquist2010discipline}. Iglesias and Pecharrom \cite{iglesias2007scaling}  showed the varying ranges of the h-index across fields and suggested a multiplicative correction to the h-index based on the scientific field to compare the scientists' research impact from different areas.

\textbf{Journal quality:} Reputable journals increase the visibility of papers and the probability of receiving citations. Petersen and Penner \cite{petersen2014inequality} found that publishing in high-quality journals decreases the average time interval between the author's future publications in those journals and has a cumulative citation advantage for the author. 

\textbf{Open access:} Free access to publications in online form increases the probability of reading and citing papers. Various studies investigated the Open Access Citation Advantage (OACA), and most found a positive effect on received citations \cite{xie2022open,blair2020open,ottaviani2016post,amjad2022investigating}. Langham-Putrow et al. \cite{langham2021open} did a systematic review of the OACA and reported that among 143 studies, 47.8\% confirmed OACA, 37\% found no OACA, and 24\% found OACA for a subset of their sample. Also, the result of our recent study  \cite{fraser2020relationship} showed substantially higher citations for preprint papers, making publications freely available. Momeni et al. \cite{momeni2022factors} examined the association of open access publishing with received citations and found a higher percentage of highly cited papers published in the open-access model than those in the closed-access model. 

\subsubsection{Coauthor Factors}
The number of the paper's citations received reveals the scientific impact of all authors, and hence it can vary according to their collaboration pattern. Hsu and Huang \cite{hsu2011correlation} found a positive correlation between the number of coauthors and received citations. Also, the result of the study by Puuska et al. \cite{puuska2014international} showed fewer citation scores for single-authored publications. Sarig{\"o}l et al. \cite{sarigol2014predicting} tried to predict highly cited papers via the centrality of their authors in the co-authorship network and found a positive correlation between highly cited publications and highly centralized authors. 

Other studies \cite{puuska2014international, ni2018relationship} examined the citation impact of international coauthors and demonstrated a positive relation between international collaboration and received citations. 
\subsection{Prediction Approaches}
Many studies employed machine learning regression and classification approaches to predict the scientific impact of publications and researchers \cite{weihs2017learning, nikolentzos2021can, abrishami2019predicting, ruan2020predicting, jiang2021hints, wu2019predicting}. The most common methods in these studies were regression models such as Support Vector Regression (SVR), Gradient Boosted Regression Trees (GBRT) or Gradient Boosting (GB), Gradient-Boosting Decision Tree (GBDT), Extreme Gradient Boosting (XGBoost), Random Forest (RF), K-nearest Neighbour (KNN), and Neural Networks (NN). Nie et al. \cite{nie2019academic} introduced a classification method to detect the academic rising stars (who have a lower research profile at the beginning of their career compared to other authors in the same career stage but may become prominent contributors in the future) and found better performance for KNN algorithm for small datasets, but a relatively stable result for GBDT, GB, RF, and RF with the change of dataset size. Ruan et al. \cite{ruan2020predicting} examined the performance of different regression algorithms and reported the best performance for Backpropagation neural network. Wu et al. \cite{wu2019predicting} examined SVR, RF, GBRT, and XGBoost regression models for h-index prediction and obtained the best performance for XGBoost. 
The performance of methods for predicting the h-index in different ranges depends on applied features. By using prior impact-based features and regression models, previous studies \cite{wu2019predicting,ayaz2018predicting,weihs2017learning} presented models with ${R^2>0.90}$ for the first predicting year and decreased in the next predicting years. However, none of these studies investigated the extent of the contribution of different features in the prediction task. Our study examines the contribution of features to the h-index prediction via feature selection/ranking approaches to understanding the influential factors better.

\section{Data and Methods}\label{secMD}
\subsection{Describing the Dataset }\label{subsecDD}
We used the in-house Scopus database maintained by the German Competence Centre for Bibliometrics (Scopus-KB), 2020 version, as the central resource of analyses and employed Scopus author Id to identify authors. We defined the career age of authors by the years between the first and last publication time. We took authors who started publishing after 1994 and used their publications until 2008 to calculate the features' value. We detected the gender status of authors by a combined name and image-based approach introduced by Karimi et al. \cite{karimi2016inferring}, which results in a binary variable. We acknowledge that a person's gender can not be split into male and female, and if we consider the social dimensions, we have more gender identities.

To remove ``not active authors'' from the analyzed data, we included just those authors who had at least five years of career age, an h-index higher than zero and matched the threshold of one publication per three years in their career age. Excluding authors without gender status results in a final list of 1,824,203 authors. Table \ref{TableStatisticAuthorPaper} presents some information about the distribution of analysed papers among main research domains (categorized by the All Science Journal Classification (ASJC) System in Scopus), the distribution of authors among gender, and career stages.   
\begin{table}[!ht]
\centering
 \caption{The number of analyzed papers across scientific fields and gender and career stage distribution of authors.}
 \label{TableStatisticAuthorPaper}
 \begin{adjustbox}{width={0.60\textwidth},totalheight={\textheight},keepaspectratio}%
 \begin{tabular}{  |l|c|c|} 
 \hline
 &\textbf{Number}&\textbf{Percentage}\\
  \hline
 \textbf{Papers}& 40,352,318&\\
 \hspace{0.4cm}  Health Sciences& 10,608,222 & 26.3 $\%$\\
 \hspace{0.4cm} Life Sciences& 8,831,499& 21.9 $\%$\\
 \hspace{0.4cm} Physical Sciences& 17,089,343 &42.3 $\%$\\
 \hspace{0.4cm} Social Sciences $\&$ Humanities& 3,272,508&8.1$\%$\\
 \hspace{0.4cm} Multidisciplinary& 550,746 & 1.4$\%$\\
 \hline
\textbf{Authors} & 1,824,203 & \\
\hspace{0.2cm}Gender:& & \\
 \hspace{0.4cm}Female& 543,517 & 30$\%$\\
 \hspace{0.4cm}Male & 1,280,686&70$\%$\\
 \hspace{0.2cm}Career stage:& & \\
 \hspace{0.4cm}Junior & 265,368&15$\%$\\
  \hspace{0.4cm}mid-level & 533,768&29$\%$\\
 \hspace{0.4cm}senior & 1,025,067&56$\%$\\

\hline
 \end{tabular}
 \end{adjustbox}
\end{table}

We applied the prediction model to three datasets containing the authors regarding their career development:
\begin{itemize}
   \item Junior: researchers with a career age of fewer than five years (the first publication between 2005 and 2008)
    \item Mid-level: researchers with a career age between 5 and 9 years (the first publication between 2000 and 2004)
    \item Senior: researchers with a career age of over ten years (the first publication between 1995 and 1999).
\end{itemize}

\subsection{Feature Engineering}\label{subsecMD}

Table \ref{TableFeatures} shows variables used to estimate the future h-index of researchers. In this table, we mentioned the previous studies that employed any of the features for the prediction task. In the following, we explain how we calculated the features:

\begin{itemize}

\item \textbf{\textit{Gender:}} It has a value equal to one for males and zero for females.

\item \textbf{\textit{MobilityScore:}} This feature indicates the frequency of movement between countries by tracking the authors' affiliations over their publications. More details about calculating this feature are available in our previous study \cite{momeni2022many}.   

\item \textbf{\textit{IncomeCurrentCountry:}} This feature indicates the countries' income level based on the GDP per capita of the affiliation country in the last publication.  We used the World Bank information\footnote{\href{https://www.weforum.org/agenda/2020/08/world-bank-2020-classifications-low-high-income-countries/}{https://www.weforum.org/agenda/2020/08/world-bank-2020-classifications-low-high-income-countries/}}  to measure it.

\item \textbf{\textit{PrimaryAuthorRatio:}} We defined the primary author as the first or corresponding author. We computed the value of this feature by dividing the number of publications in which the researcher is the primary author to all publications.

\item \textbf{\textit{OpenAccessRatio:}} We extracted the article's access status from the Unpaywall dataset (a service that provides full-text articles from open access resources\footnote{\href{https://unpaywall.org/}{https://unpaywall.org/}}). An open-access article can be any form of gold, green, or bronze. We declare that we could match from 8,953,939 investigated papers only 5,476,852 (61\%) with Unpaywall's articles. To calculate the proportion of open access papers, we considered the number of detected as open access to the number of whole articles of the author. 

\item \textbf{\textit{MainField:}} We identified the field of authors from the field of the journals in which they publish, and in Scopus are classified under four broad subject clusters\footnote{\href{https://service.elsevier.com/app/answers/detail/a_id/14882/supporthub/scopus/~/what-are-the-most-frequent-subject-area-categories-and-classifications-used-in/}{Subject Area}}. The field with the most publications will be the main field of the author.

\item \textbf{\textit{HighRankPapersRatio:}} We used the journal ranking based on their quality to evaluate the rank of papers.
To assess the quality of journals, we calculated the h-index of journals from 1995 to 2015. Because of different citation patterns among disciplines, journals' h-index can have varying ranges for different disciplines, which should be normalized. We applied the percentile rank approach inspired by Bornmann and Lutz \cite{bornmann2014p100} and computed the h-index's rank among all journals inside its discipline. We used Scopus's classification system to find the journals' disciplines. In this system, journals are classified into 27 subject categories\footnote{\href{https://service.elsevier.com/app/answers/detail/a_id/14882/supporthub/scopus/~/what-are-the-most-frequent-subject-area-categories-and-classifications-used-in/}{Subject Area Classifications}}. In this percentile rank approach, each journal within a category ranks 0 (lowest h-index) to 100 (highest h-index). Journals with the same h-index have the same rank. If the journal belongs to more than one category, we used the weighted Percentile Ranking wPR) \cite{bornmann2020evaluation}. Based on this approach, wPR will be calculated using the formula:

\begin{equation}
\begin{split}
 wPR = \frac{PR_{sc1} * n_{sc1} +PR_{sc2} * n_{sc2} +...+PR_{sci} * n_{sci}}{n_sc1 +n_sc2 +...+n_{sci}}
 \end{split}
 \label{equNorm}
\end{equation}

Whereby $sci$ is the  \textit{i}th subject category that the journal belongs to and $n_{sci}$ is the number of journals in this subject category, and $PR_{sci}$ is PR of the journal in it. Journals with a wPR higher than 50\% are assumed to be high quality. Finally, we counted the proportion of the author's publications in high-quality journals among all their publications for the variable \textit{HighRankPapersRatio}.

\item \textbf{\textit{DisciplineMobility:}} This feature indicates the number of unique fields the author has published during the entire academic age divided by the number of whole papers.

\item \textbf{\textit{KeywordPopularity:}} This feature indicates the proportion of papers with popular keywords. First, we ranked keywords based on the frequency of occurrence in papers from the same discipline (27 subject categories) and publication year to measure the keyword popularity for a paper. Next, we gave a value of 1 to the paper with a ranking above 0.5; otherwise, 0. Finally, we summed up these values over all papers and divided them by the number of all papers.

\item \textbf{\textit{EnglishPapersRatio:}} This feature measures the ratio of papers written in English.

\item \textbf{\textit{CoauthorPerPaper:}} This feature displays the number of unique coauthors, which is normalized by dividing by the number of all papers.

\item \textbf{\textit{CoauthorMaxHindex:}} To assess the effect of the scientific impact of coauthors, we used the maximum h-index among all coauthors as an alternative measure of the Godfather Effect \cite{mccarty2013predicting}.  

\item \textbf{\textit{InternationalCoauthorRatio:}} This feature specifies the number of international collaborators for all papers. To calculate it, first, we counted the number of papers with at least one coauthor having a different country in the affiliation than the author and then divided it by the number of all papers.

\end{itemize}

\begin{table}[!ht]
\centering
 \caption{Features used to train the machine learning models to predict the h-index.}
 \label{TableFeatures}
 \begin{adjustbox}{width={\textwidth},totalheight={\textheight},keepaspectratio}%
 \begin{tabular}{  |c|l|p{0.5\textwidth}|l| } 
 \hline
 \textbf{Feature group} & \textbf{Feature name} & \textbf{Description} & \textbf{Studies}\\ 
 \hline
 \multirow{5}{*}{Demographic}  & \textit{CareerAge} & Years since first publication & \cite{ayaz2018predicting} \\ 
   & \textit{Gender} & Zero for females and one for males &\\
  & \textit{MobilityScore} & Number of changing the affiliation at the country level &\\
  & \textit{IncomeCurrentCountry} & GDP Per Capita of current affiliation country &\\
 \hline
  \multirow{4}{*}{Prior Impact} & \textit{CurrentHindex} & Current h-index & \cite{ayaz2018predicting}; \cite{weihs2017learning}; \cite{wu2019predicting}\\ 
& \textit{PaperPerYear} & Number of total papers divided by career age & \cite{ayaz2018predicting}; \cite{weihs2017learning}; \cite{wu2019predicting}\\ 
 & \textit{CitationPerPaper} &  Number of total citations among all papers until 2008 divided by the number of all papers & \cite{ayaz2018predicting}; \cite{weihs2017learning}; \cite{wu2019predicting}\\
 \hline
  \multirow{14}{*}{Paper/Venue} & \textit{PrimaryAuthorRatio} & Number of papers being as primary author divided by the number of all papers &\\
 & \textit{OpenAccessRatio} & Number of open access papers divided by the number of all papers &\\
 & \textit{MainField} & The scientific field with the highest amount of publications & \\
 & \textit{HighRankPapersRatio} & Number of publications in high-quality journals divided by the number of all papers & \cite{ayaz2018predicting}\\
 & \textit{DisciplineMobility} & Number of unique disciplines authors has published paper divided by the number of all papers &\\
 & \textit{KeywordPopularity} & Number of publications with at least one popular keyword divided by the number of all papers &\\
 & \textit{EnglishPapersRatio} & Number of English papers divided by the number of all papers &\\
 \hline
  \multirow{4}{*}{Coauthor}  & \textit{MaxCoauthorHindex} & Maximum h-index of coauthors among all papers & \cite{mccarty2013predicting} 
  \\ 
  & \textit{CoauthorPerPaper} & Number of unique coauthors among all publications divided by the number of all papers & \cite{wu2019predicting}\\
& \textit{InternationalCoauthorRatio} & Number of papers with international collaboration divided by the number of all papers & \\
 \hline
 \end{tabular}
 \end{adjustbox}
\end{table}

We provided descriptive statistics for investigated features in Table \ref{TableFeatureDescription} to describe the data.

\begin{table}[!ht]
\centering
 \caption{Descriptive statistics of features. This table shows the mean standard deviation for numerical features and distribution of authors based on their gender, mobility status and main field.}
 \label{TableFeatureDescription}
 \begin{adjustbox}{width={\textwidth},totalheight={\textheight},keepaspectratio}%
 \begin{tabular}{  |l|c|c|l| } 
 \hline
 \textbf{Feature name} & \textbf{Mean} & \textbf{Standard deviation} & \textbf{Distribution}\\ 
 \hline
\textit{CareerAge} & 9.35 & 3.69 & \\
\textit{Gender} & 0.70 & 0.46 & 70$\%$ male, 30$\%$ female \\
\textit{MobilityScore} & 0.50 & 1.08 & 27$\%$ mobile, 73$\%$ non-mobile \\
\textit{IncomeCurrentCountry} & 35,052.63 & 14,024.40 & \\
\hline
\textit{CurrentHindex} & 6.13 & 6.17 & \\
\textit{PaperPerYear} & 2.00 & 2.39 & \\
\textit{CitationPerPaper} & 11.47 & 22.18 & \\
\hline
\textit{PrimaryAuthorRatio} & 0.36 & 0.29 & \\
\textit{OpenAccessRatio} & 0.19 & 0.23 & \\
\textit{MainField} & & & H: 29$\%$, L:23$\%$, P:37$\%$, S:6$\%$, M:4$\%$ * \\
\textit{HighRankPapersRatio} & 0.01 & 0.06 & \\
\textit{DisciplineMobility} & 0.47 & 0.45 & \\
\textit{KeywordPopularity} & 0.53 & 0.28 & \\
\textit{EnglishPapersRatio} & 0.92 & 0.20 & \\
\hline
\textit{MaxCoauthorHindex} & 15.51 & 14.86 & \\
\textit{CoauthorPerPaper} & 3.74 & 30.39 & \\
\textit{InternationalCoauthorRatio} & 0.21 & 0.25 & \\
\hline
\multicolumn{4}{l}{* H: Health Sciences, L: Life Sciences, P: Physical Sciences, M:Multiple Fields} \\
 \end{tabular}
 \end{adjustbox}
\end{table}

\subsection{Applied Methods for the Prediction Task}\label{subsecAMPT}
We tackled the h-index prediction as a regression problem comparable to previous studies \cite{ayaz2018predicting,dong2016can,ruan2020predicting,weihs2017learning,wu2019predicting}.
We explored the performance of four different machine learning methods, namely SVR, RF, GB, and XGBoost. Among these, XGBoost emerged as the top-performing method, consistent with the findings reported by \cite{wu2019predicting}. Consequently, we utilized the XGBoost approach for our h-index prediction task.
XGBoost is a scalable end-to-end tree boosting system introduced by Chen and Guestrin \cite{chen2016xgboost}. It efficiently implements Gradient Boosting in terms of speed and is appropriate for solving problems using  minimal resources. We need to have the data in numerical form to apply this method. We utilized one hot encoder to convert the categorical values to integers. In this encoding method, each value of the categorical variable will be converted to a feature with a binary value, where 1 represents the data value and 0 is used for all other values. So, for \textit{MainField} with five values, we have five features, and the feature with a value equal to 1 indicates the \textit{MainField}. 
To evaluate the model, we utilized the Mean Absolute Percentage Error (MAPE) to measure the error as a percentage, which is appropriate to compare the performance of a model for the different datasets, as used by some previous studies \cite{bai2019predicting,wu2019predicting,weihs2017learning}. Because MAPE is affected by outliers \cite{blasco2013using}, we also utilized symmetric Mean Absolute Percentage Error (sMAPE), which is scaled to percentage too and is more resistant to outliers \cite{chen2016xgboost}. In addition, we used Root Mean Square Error (RMSE) to evaluate the performance of models, as in prior works \cite{ abrishami2019predicting,ayaz2018predicting,bai2019predicting}.
We used the k-fold cross-validation procedure to evaluate the models and fixed the k to 5. 

We defined different feature combinations based on the attributes of the author, paper, venue, and coauthors to see which feature categories are better for short/long-term prediction.
Table \ref{TableFeatCombi} shows the different combination sets utilized to train the model.

\begin{table}[ht!]
\centering
 \caption{Different feature combinations to predict the h-index.}
 \label{TableFeatCombi}
 \begin{adjustbox}{width={\textwidth},totalheight={\textheight},keepaspectratio}%
 \begin{tabular}{  |cl|l|l|l|l|l|l|l|l|l| } 
 \cline{3-11}
 \multicolumn{2}{c|}{}& \multicolumn{9}{c|}{\textbf{Feature combination}}\\
 \hline
\textbf{Feature group} & \textbf{Feature name} & 1 & 2&3&4&5&6&7&8&9\\
 \hline
  \multirow{4}{*}{\textit{Demographic}} & \textit{CareerAge}  & & \checkmark& & \checkmark& & &\checkmark & &\checkmark \\
  & \textit{Gender}   & & \checkmark& & \checkmark& & &\checkmark & &\checkmark \\
  & \textit{MobilityScore} &  & \checkmark & &\checkmark & & & \checkmark & &\checkmark \\
  & \textit{IncomeCurrentCountry}  & &\checkmark & & \checkmark& & & \checkmark & & \checkmark\\
 \hline
  \multirow{2}{*}{\textit{Prior impact}} & \textit{CurrentHindex}   &\checkmark  &   \checkmark  & \checkmark&\checkmark & \checkmark& & & & \\
   & \textit{CitationPerPaper}  &\checkmark  & \checkmark & \checkmark&\checkmark &\checkmark& & & &\\
   \hline
 \multirow{7}{*}{\textit{Paper/venue}}& \textit{PrimaryAuthorRatio}  &\checkmark  &\checkmark  &\checkmark &\checkmark& & \checkmark& \checkmark &\checkmark & \checkmark\\
 & \textit{OpenAccessRatio}  &\checkmark  &\checkmark  &\checkmark &\checkmark& &\checkmark &\checkmark&\checkmark &\checkmark \\
 & \textit{MainField}  &\checkmark  &\checkmark  &\checkmark &\checkmark & & \checkmark &\checkmark & \checkmark& \checkmark\\
 & \textit{HighRankPapersRatio}  &\checkmark  &\checkmark &  \checkmark&\checkmark & &\checkmark & \checkmark&\checkmark&\checkmark \\
 & \textit{DisciplineMobility}  &\checkmark &\checkmark &\checkmark &\checkmark & &\checkmark &\checkmark&\checkmark & \checkmark\\
 & \textit{EnglishPapersRatio}  &\checkmark &\checkmark &\checkmark &\checkmark & &\checkmark &\checkmark&\checkmark & \checkmark\\
 & \textit{KeywordPopularity} &\checkmark &\checkmark &\checkmark &\checkmark & &\checkmark &\checkmark&\checkmark & \checkmark\\
 \hline
  \multirow{3}{*}{\textit{Coauthor}}  & \textit{MaxCoauthorHindex} & & & \checkmark& \checkmark& & & &\checkmark & \checkmark \\ 
  & \textit{CoauthorPerPaper} &  & &\checkmark &\checkmark& & & &\checkmark &\checkmark\\
  & \textit{InternationalCoauthorRatio} &  & &\checkmark &\checkmark& & & &\checkmark &\checkmark\\
\hline
 \end{tabular}
 \end{adjustbox}
\end{table}

\subsection{Temporal Dimension of Feature Performance}
Prior studies regarded varying time frames to estimate the future h-index \cite{dong2015will,ayaz2018predicting,wu2019predicting} and examined several years from one to five-year and \cite{dong2015will} for five-year and ten-year time frames. The prediction performance declined as the prediction time frame increased in all studies. We considered the h-index as our target from one to ten years in the future (h-index from 2009 to 2018). It enables us to measure the extent of predicting performance in the future.

\subsection{Feature Impact Analysis}
To examine the importance of each feature in the prediction task, we employed a feature selection technique,  \textit{Recursive Feature Elimination} (RFE), which removes recursively features and builds a model based on the remaining features \cite{artur2021review, zhao2022rfe}. 

\section{Results}\label{secR}
In this section, we present the results of our analysis, focusing on the relationship between various features and the future h-index of researchers. Before delving into the specific findings, we address the potential multicollinearity problem in Section \ref{result_corr} by examining the dependencies between features. We analyze the Pearson correlation between independent variables and visualize the results using a heatmap. Next, we explore the correlation between the introduced features and the future h-index in 2009, 2014, and 2018. This analysis allows us to examine the statistical association between variables, providing insights into the strength and direction of these relationships. However, it's important to note that the correlations captured by the correlation analysis primarily represent linear associations between features and the h-index.

To capture the non-linear relationship between the h-index and the investigated features, we apply ML prediction models in Section \ref{result_predict}. First, in Section \ref{result_importance}, we identify the most important factors for predicting the h-index using the feature selection method, RFE. This step helps us narrow down the key variables. Then, in Section \ref{result_temporal}, we examine the effectiveness of these models for researchers with different career ages, focusing on the temporal dimension.
\subsection{Correlation Analysis}\label{result_corr}
\begin{figure}
    \includegraphics[width=0.96\linewidth]{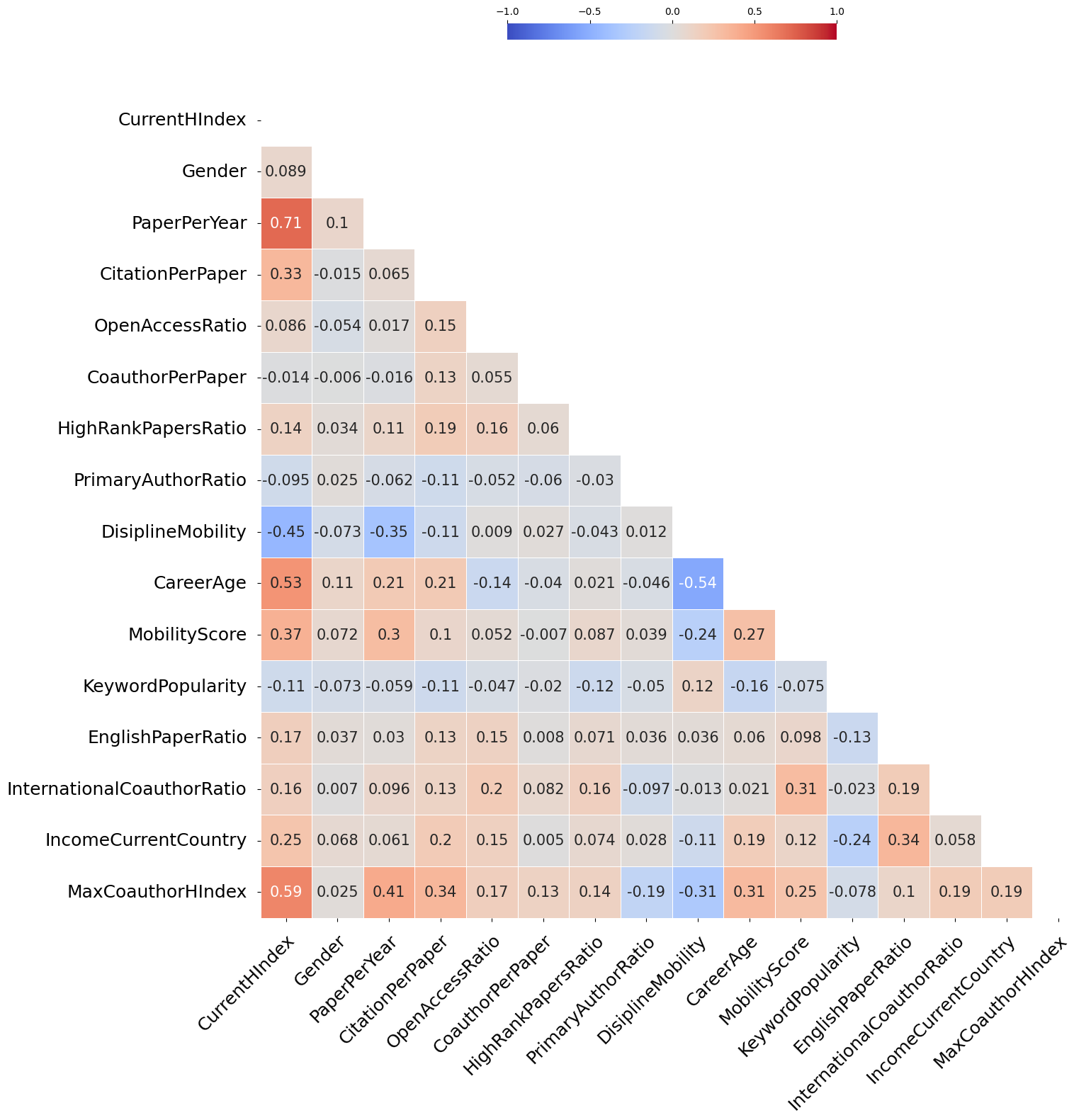}
    \caption{Pearson correlation between independent variables }
    \label{fig:heatmap}
\end{figure}

Before investigating the relationship between various features and future h-index, we examine the dependencies between features to avoid the potential multicollinearity problem. Figure \ref{fig:heatmap} presents the Pearson correlation between independent variables. We see a strong correlation between \textit{PaperPerYear} and \textit{CurrentHindex}; therefore, to avoid multicollinearity in regression and classification models, we exclude \textit{PaperPerYear} from the data for prediction tasks.

To examine the affecting factors on the h-index, we first provide the correlation between features introduced in Table \ref{TableFeatures} and future h-index. Table \ref{table_corr} presents the Pearson correlation coefficient between the features (except for \textit{MainField}, a categorical variable) and h-index in 2009, 2014, and 2018. The highest correlation coefficient for two prior impact-based features (\textit{CurrentHindex}, \textit{PaperPerYear}) displays the strong association of this kind of feature with the future h-index. The higher correlation coefficient between the future h-index and the number of papers (\textit{PaperPerYear}) than the number of citations (\textit{CitationPerPaper}) reveals that productivity has a more significant impact than received citations on the h-index. Among non-prior impact-based features, \textit{MaxCoauthorHindex} has the highest correlation with the h-index and suggests the strong relation of coauthors' reputation with the future h-index. The negative value for \textit{DisciplineMobility} suggests that authors who publish in several scientific fields have a lower h-index than those who publish in a specific field.

Most of the correlations between the influential factors and the h-index demonstrate consistent patterns across different time frames, indicating similar effects in both the short and long term. While correlation analysis offers informative perspectives about the strength and direction of these relationships, it primarily captures linear associations between variables. However, we will employ machine learning algorithms in the next section to uncover non-linear associations and delve deeper into the temporal dimension of the relationship for researchers in different career stages. This approach allows us to examine the complex interactions and temporal dynamics between the factors and the h-index, specifically analyzing how they vary across different career stages.  It provides a more comprehensive understanding of their relationship and enables us to make accurate predictions beyond what correlation analysis alone can reveal.

\begin{table}
\centering
\caption{Pearson correlation coefficient between the features and h-index in the future for three different years. \textit{CurrentHindex}, \textit{PaperPerYear}, and \textit{CitationPerPaper} are prior impact-based features, and the rest are non-prior impact-based features.}
\begin{tabular}{ |l|c|c|c|c|}
\cline{2-4}
\multicolumn{1}{c|}{}&\multicolumn{3}{c|}{\textbf{H-index}}\\
\hline
\textbf{Feature} & \textbf{2009} & \textbf{2014} & \textbf{2018} \\
\hline
\hspace{0.2cm}\textit{CareerAge} &0.48&0.38&0.32\\
\hspace{0.2cm}\textit{Gender}&0.09&0.08&0.07\\
\hspace{0.2cm}\textit{MobilityScore}&0.44&0.43&0.41\\
\hspace{0.2cm}\textit{IncomeCurrentCountry}&0.23&0.21&0.19\\
\hline
\hspace{0.2cm}\textit{CurrentHindex}&0.99&0.95&0.87\\
\hspace{0.2cm}\textit{PaperPerYear} &0.73&0.75&0.73\\
\hspace{0.2cm}\textit{CitationPerPaper}&0.31&0.26&0.23\\
\hline
\hspace{0.2cm}\textit{PrimaryAuthorRatio} &-0.09&-0.08&0.-0.06\\
\hspace{0.2cm}\textit{OpenAccessRatio} &0.10&0.14&0.15\\
\hspace{0.2cm}\textit{EnglishPapersRatio}&0.17&0.16&0.15\\
\hspace{0.2cm}\textit{KeywordPopularity}&-0.09&-0.07&-0.05\\
\hspace{0.2cm}\textit{HighRankPapersRatio}&0.14&0.15&0.15\\
\hspace{0.2cm}\textit{DisciplineMobility} &-0.45&-0.42&-0.39\\
\hline
\hspace{0.2cm}\textit{MaxCoauthorHindex} &0.58&0.58&0.55\\
\hspace{0.2cm}\textit{CoauthorPerPaper} &-0.01&0.02&0.04\\
\hspace{0.2cm}\textit{InternationalCoauthorRatio} &0.17&0.19&0.19\\
\hline
\end{tabular}
\label{table_corr}
\end{table}

\subsection{Prediction Analysis}\label{result_predict}
In this section, we present the prediction results of our study, highlighting the influence of different features on predicting the h-index. Firstly, in Section \ref{result_importance}, we evaluate the importance of these features using the Recursive Feature Elimination (RFE) method. Then, in Section \ref{result_temporal}, we examine the effectiveness and stability of various feature combinations in predicting the h-index. We analyze the predictive performance across different time frames and for researchers at different career stages, providing valuable insights into the temporal dynamics and the impact of features on the h-index prediction task.

\subsubsection{Feature Impact}\label{result_importance}
We evaluate the importance of features in the prediction task by ranking them via the RFE method. Table \ref{tableFeatureSelectRFE} demonstrates the feature ranking for selecting the predictors in the model. For \textit{MainField}, we used one hot encoder, which converts each unique category value to a feature (five features for five fields). The features highlighted in blue are the top five features in the selection process. We observe that paper-specific features are most relevant among all career stages. Also, coauthor-specific features are among the most important features to predict the h-index for the researchers in junior and mid-level career stages. It suggests that the coauthor's characteristics have more influence on the h-index for these researchers than seniors.

\begin{table}[!ht]
\centering
\caption{Ranking of features for selection in predicting the h-index with the RFE method. The five most relevant features (with a ranking between 1 and 5) are highlighted in blue.}
\label{tableFeatureSelectRFE}
\begin{adjustbox}{width={\textwidth},totalheight={\textheight},keepaspectratio}%
\begin{tabular}{  |l|ccc|ccc|ccc|} 
\hline
\textbf{Career stage} &\multicolumn{3}{c|}{\textbf{Junior}}&\multicolumn{3}{c|}{\textbf{Mid-level}}&\multicolumn{3}{c|}{\textbf{Senior}}\\
\hline
\textbf{Prediction year}&\textbf{2009}&\textbf{2014}&\textbf{2018}&\textbf{2009}&\textbf{2014}&\textbf{2018}&\textbf{2009}&\textbf{2014}&\textbf{2018}\\
\hline
\textbf{Feature:}&Rank&Rank&Rank&Rank&Rank&Rank&Rank&Rank&Rank\\
\hspace{0.2cm}\textit{CareerAge} &7&\colorbox{cyan}{5}&\colorbox{cyan}{4}&\colorbox{cyan}{3}&\colorbox{cyan}{2}&\colorbox{cyan}{3}&\colorbox{cyan}{3}&\colorbox{cyan}{4}&\colorbox{cyan}{2}\\ 
\hspace{0.2cm}\textit{Gender} &20&18&16&19&18&16&19&19&19\\ 
\hspace{0.2cm}\textit{MobilityScore} &18&14&12&12&8&\colorbox{cyan}{4}&18&18&16\\ 
\hspace{0.2cm}\textit{IncomeCurrentCountry} &14&16&17&13&14&13&9&9&9\\ \hline
\hspace{0.2cm}\textit{CurrentHindex}&\colorbox{cyan}{1}&\colorbox{cyan}{1}&\colorbox{cyan}{1}&\colorbox{cyan}{1}&\colorbox{cyan}{1}&\colorbox{cyan}{1}&\colorbox{cyan}{1}&\colorbox{cyan}{1}&\colorbox{cyan}{1}\\ 
\hspace{0.2cm}\textit{CitationPerPaper}&11&15&15&6&6&7&7&\colorbox{cyan}{5}&6\\ 
\hspace{0.2cm}\textit{PrimaryAuthorRatio} &6&10&9&10&9&9&16&11&10\\ 
\hspace{0.2cm}\textit{OpenAccessRatio} &\colorbox{cyan}{3}&8&7&\colorbox{cyan}{4}&7&8&6&\colorbox{cyan}{2}&\colorbox{cyan}{5}\\ 
\hspace{0.2cm}\textit{EnglishPapersRatio}&\colorbox{cyan}{4}&11&13&9&16&17&15&17&18\\ 
\hspace{0.2cm}\textit{KeywordPopularity}&10&13&14&11&13&15&11&14&13\\ 
\hspace{0.2cm}\textit{MainField} &&&&&&&&&\\ 
\hspace{0.4cm}\textit{Health Sciences}&12&\colorbox{cyan}{3}&\colorbox{cyan}{5}&17&19&18&\colorbox{cyan}{5}&15&17\\ 
\hspace{0.4cm}\textit{LifeSciences}&15&17&18&15&17&19&8&6&\colorbox{cyan}{3}\\ 
\hspace{0.4cm}\textit{multiple fields}&19&20&20&20&20&20&20&20&20\\ 
\hspace{0.4cm}\textit{Physical Sciences}&13&\colorbox{cyan}{2}&\colorbox{cyan}{2}&16&\colorbox{cyan}{4}&\colorbox{cyan}{5}&13&7&8\\ 
\hspace{0.4cm}\textit{Social Sciences}&16&19&19&14&15&14&\colorbox{cyan}{4}&\colorbox{cyan}{3}&\colorbox{cyan}{4}\\ 
\hspace{0.2cm}\textit{HighRankPapersRatio}&9&9&11&\colorbox{cyan}{5}&12&12&12&16&15\\ 
\hspace{0.2cm}\textit{DisciplineMobility}  &\colorbox{cyan}{2}&12&10&\colorbox{cyan}{2}&11&11&\colorbox{cyan}{2}&12&14\\ \hline
\hspace{0.2cm}\textit{MaxCoauthorHindex} &\colorbox{cyan}{5}&6&\colorbox{cyan}{3}&8&\colorbox{cyan}{5}&6&10&10&11\\ 
\hspace{0.2cm}\textit{CoauthorPerPaper}&17&\colorbox{cyan}{4}&6&18&10&10&17&8&7\\ 
\hspace{0.2cm}\textit{InternationalCoauthorRatio}  &8&7&8&7&\colorbox{cyan}{3}&\colorbox{cyan}{2}&14&13&12\\ \hline
\end{tabular}
\end{adjustbox}
\end{table}

\subsubsection{Career Stage and Temporal Dimension of Model Performance}\label{result_temporal}
\begin{figure}[ht]
    \centering
    \includegraphics[width=0.95\linewidth]{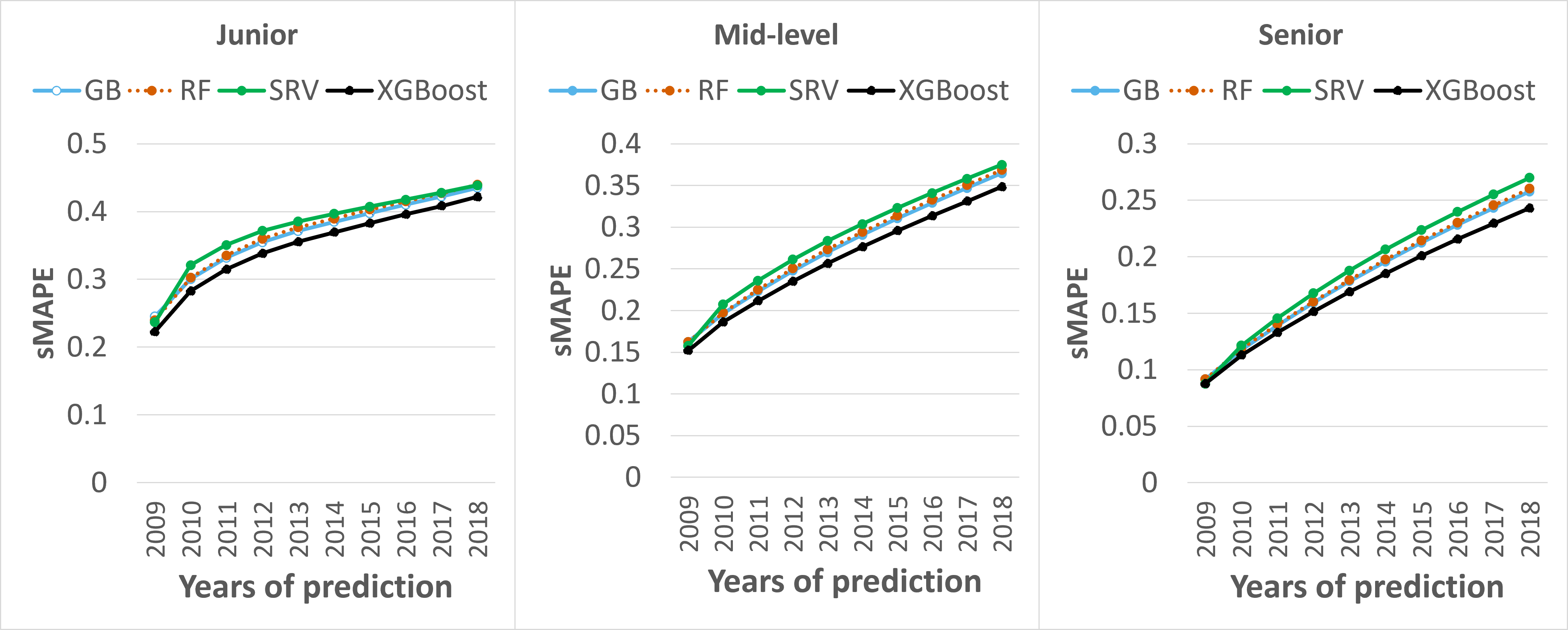} 
    \caption{Comparison of predictive performance using sMAPE metric among four machine learning algorithms (SVR, RF, GB, and XGBoost) for researchers' h-index prediction at different career stages from 2009 to 2018. The analysis utilized feature set 1 as the predictor.}
    \label{fig:performance_methods}
\end{figure}

Before we show the result of the analyses, we make some comparisons between the performance of our model and previous works. Wu et al. \cite{wu2019predicting} have already compared their performance with other studies \cite{dong2015will,weihs2017learning,ayaz2018predicting} and presented the best performance among all these studies. They excluded the authors with an h-index of less than four from the investigated data. They achieved the minimum MAPE of 0.063 for the first prediction year by employing more prior impact-based features. We could reach the minimum MAPE of 0.068 by applying this condition to investigated authors. Instead, two-thirds of the authors will be discarded in our analyses. Because of losing too much data, particularly from young scholars, we didn't apply this condition and implemented our models with all authors, despite reducing the performance.
To evaluate the predictive performance, we conducted a comparison among four machine learning algorithms: SVR, RF, GB, and XGBoost, using feature set 1, which includes all features. The results are illustrated in Figure \ref{fig:performance_methods}, demonstrating that XGBoost outperforms the other methods across all career stages. As a result, we proceed with this method for further analyses.

Table \ref{TablePerformanceHindex} showcases the performance metrics, including RMSE, MAPE, and sMAPE, for all three groups of researchers (junior, middle-level, and senior) across the years 2009, 2014, and 2018. It provides a detailed overview of the model's performance, enabling a direct comparison of the metrics for each group and year. Lower values of these metrics indicate better predictive performance.
We observe a decline in performance for all groups of researchers across all metrics from the near future (2009) to the far future (2018). While the models for seniors generally demonstrate better performance compared to the other groups, the decline in performance is more pronounced for researchers in later career stages.
Specifically, in terms of RMSE for junior researchers, the range varies from 0.6 (combination 4, considering all features) in 2009 to 5.46 (combination 1, considering only prior-impact features) in 2018. For seniors, the range is from 0.74 (combination 1) in 2009 to 6.93 (combination 8) in 2018. We observe a greater decline in performance for seniors in the far future compared to juniors.
When considering MAPE and sMAPE, which provide performance in percentage, we can better compare the model's performance across career stages. Although these metrics show better performance for researchers in later career stages, the performance is more stable for juniors. For instance, combination 4 exhibits the best performance for juniors, with sMAPE ranging from 0.22 to 0.42, while for seniors, it ranges from 0.09 to 0.24.
Furthermore, despite combinations containing prior-impact features exhibiting better performance in the near future (2009) for all researcher groups, we observe that for juniors, combinations without prior-impact features approach the performance of models with prior-impact features in the long term (2018). In some cases, these combinations even outperform models with prior-impact features. This finding suggests that non-prior impact-based features are more reliable predictors for the future h-index of junior researchers, compared to seniors.
In summary, seniors generally exhibit better performance, but juniors demonstrate more stable performance and the potential for improved long-term predictions using non-prior impact-based features.

\begin{table}[!htbp]
\centering
 \caption{ Comparison of XGBoost regression model performance to predict the feature h-index in one, five, and ten years (2009, 2014, and 2018) implemented on three datasets (junior, middle, and senior researchers). RMSE, MAPE, and sMAPE are the metrics to assess performance.}
 \label{TablePerformanceHindex}
 \begin{adjustbox}{width={\textwidth},totalheight={\textheight},keepaspectratio}%
 \begin{tabular}{  |c|l|l|l|l|l|l|l|l|l|l| } 
 \cline{3-11}
 \multicolumn{2}{c|}{}&  \multicolumn{3}{|c|}{\textbf{Junior}}& \multicolumn{3}{|c|}{\textbf{Middle-level}}& \multicolumn{3}{|c|}{\textbf{Senior}}\\
  \hline
   \textbf{Feature combination}&\textbf{Metric}&\textbf{2009}&\textbf{2014}&\textbf{2018}&\textbf{2009}&\textbf{2014}&\textbf{2018}&\textbf{2009}&\textbf{2014}&\textbf{2018}\\
  \hline
\multirow{3}{*}{1}&RMSE&0.62&3.01&5.15&0.68&2.85&4.94&0.75&3&5.09\\
&MAPE&0.24&0.52&0.62&0.16&0.33&0.45&0.09&0.2&0.28\\
&sMAPE&0.23&0.39&0.45&0.16&0.29&0.36&0.09&0.19&0.25\\\hline
\multirow{3}{*}{2}&RMSE&0.61&2.91&4.99&0.67&2.78&4.81&0.75&2.94&4.97\\
&MAPE&0.24&0.49&0.59&0.16&0.32&0.43&0.09&0.2&0.28\\
&sMAPE&0.23&0.38&0.43&0.15&0.28&0.35&0.09&0.19&0.25\\\hline
\multirow{3}{*}{3}&RMSE&0.61&2.85&4.91&0.68&2.75&4.77&0.75&2.9&4.9\\
&MAPE&0.24&0.5&0.6&0.16&0.33&0.44&0.09&0.2&0.28\\
&sMAPE&0.23&0.38&0.44&0.15&0.28&0.36&0.09&0.19&0.25\\\hline
\multirow{3}{*}{4}&RMSE&0.6&2.78&4.81&0.67&2.68&4.67&0.74&2.85&4.8\\
&MAPE&0.24&0.48&0.57&0.16&0.32&0.43&0.09&0.2&0.27\\
&sMAPE&0.22&0.37&0.42&0.15&0.28&0.35&0.09&0.19&0.24\\\hline
\multirow{3}{*}{5}&RMSE&0.67&3.23&5.46&0.72&3.05&5.23&0.78&3.24&5.49\\
&MAPE&0.28&0.57&0.68&0.17&0.37&0.49&0.09&0.23&0.31\\
&sMAPE&0.27&0.42&0.47&0.17&0.31&0.39&0.1&0.21&0.28\\\hline
\multirow{3}{*}{6}&RMSE&1&3.27&5.43&1.87&3.56&5.5&4.04&5.75&7.52\\
&MAPE&0.37&0.56&0.65&0.41&0.44&0.53&0.41&0.4&0.44\\
&sMAPE&0.31&0.41&0.46&0.32&0.35&0.4&0.32&0.32&0.34\\\hline
\multirow{3}{*}{7}&RMSE&0.97&3.19&5.3&1.8&3.48&5.38&3.79&5.47&7.24\\
&MAPE&0.36&0.54&0.62&0.39&0.43&0.51&0.38&0.38&0.42\\
&sMAPE&0.31&0.4&0.44&0.31&0.34&0.39&0.31&0.31&0.33\\\hline
\multirow{3}{*}{8}&RMSE&0.96&2.97&5.02&1.75&3.33&5.23&3.64&5.23&6.93\\
&MAPE&0.35&0.53&0.62&0.38&0.41&0.5&0.35&0.36&0.4\\
&sMAPE&0.3&0.4&0.44&0.31&0.33&0.38&0.29&0.29&0.32\\\hline
\multirow{3}{*}{9}&RMSE&0.94&2.92&4.93&1.69&3.28&5.15&3.47&5.05&6.74\\
&MAPE&0.34&0.51&0.6&0.36&0.41&0.49&0.34&0.34&0.39\\
&sMAPE&0.3&0.39&0.43&0.3&0.33&0.38&0.28&0.29&0.32\\
\hline
 \end{tabular}
 \end{adjustbox}
\end{table}

\begin{figure}[!ht]
\settoheight{\tempdima}{\includegraphics[width=.32\linewidth]{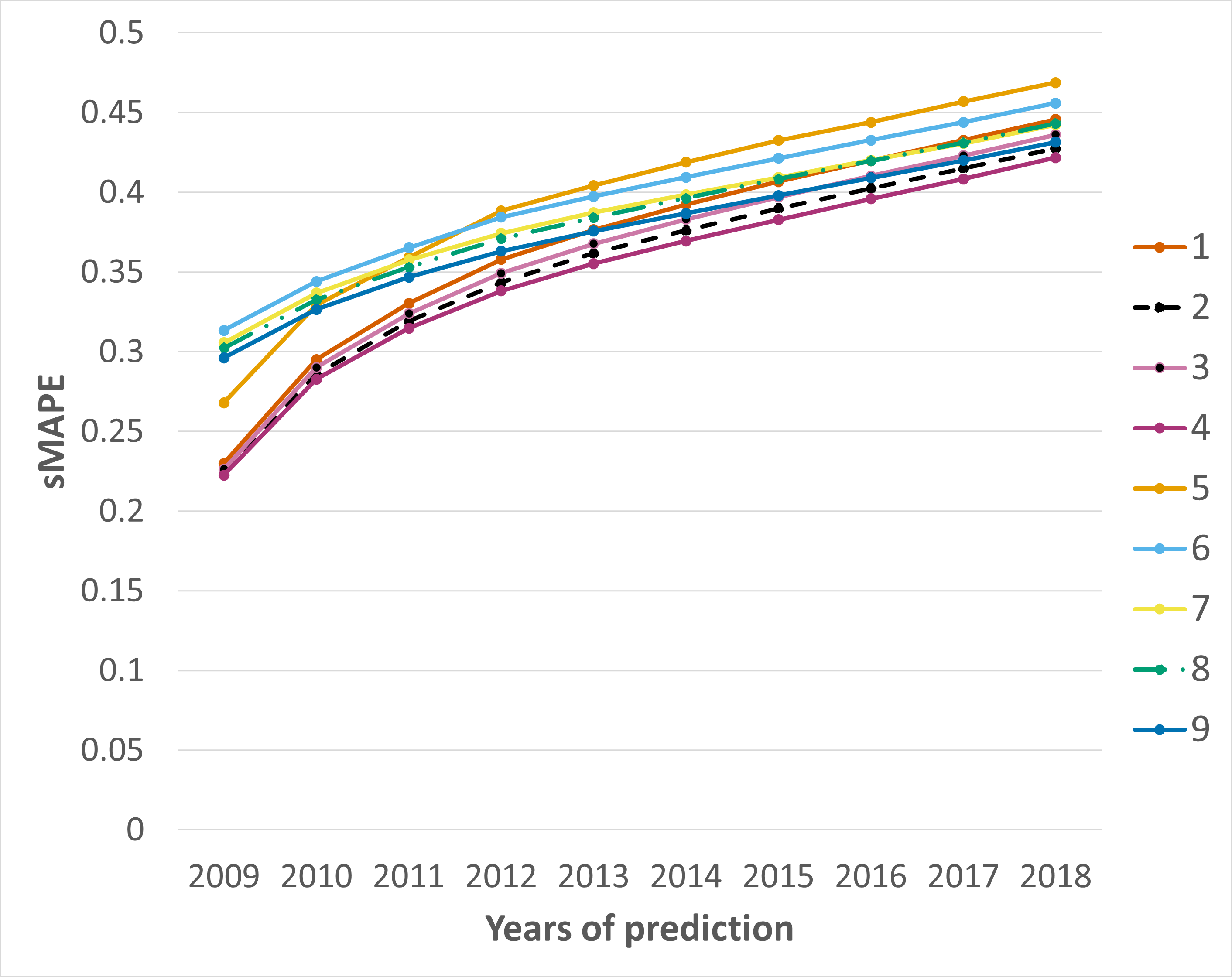}}%
\centering\begin{tabular}{@{}c@{ }c@{ }c@{ }c@{}}
&\textbf{(a)} & \textbf{(b)}  \\
\rowname{Junior}&
\includegraphics[width=.57\linewidth]{smape_juniors.png}&
\includegraphics[width=.37\linewidth]{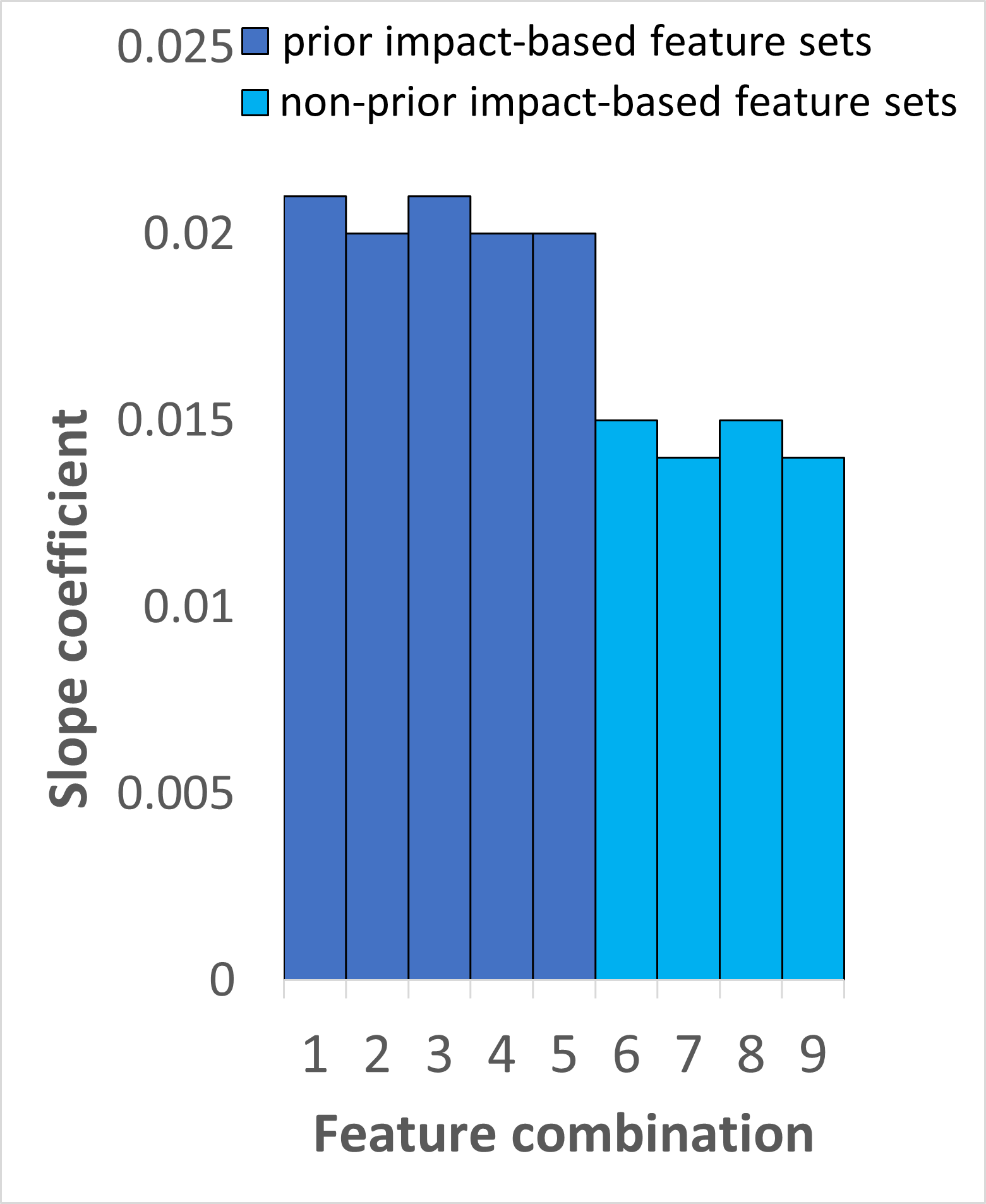}\\[-1ex]
\rowname{Mid-level}&
\includegraphics[width=.57\linewidth]{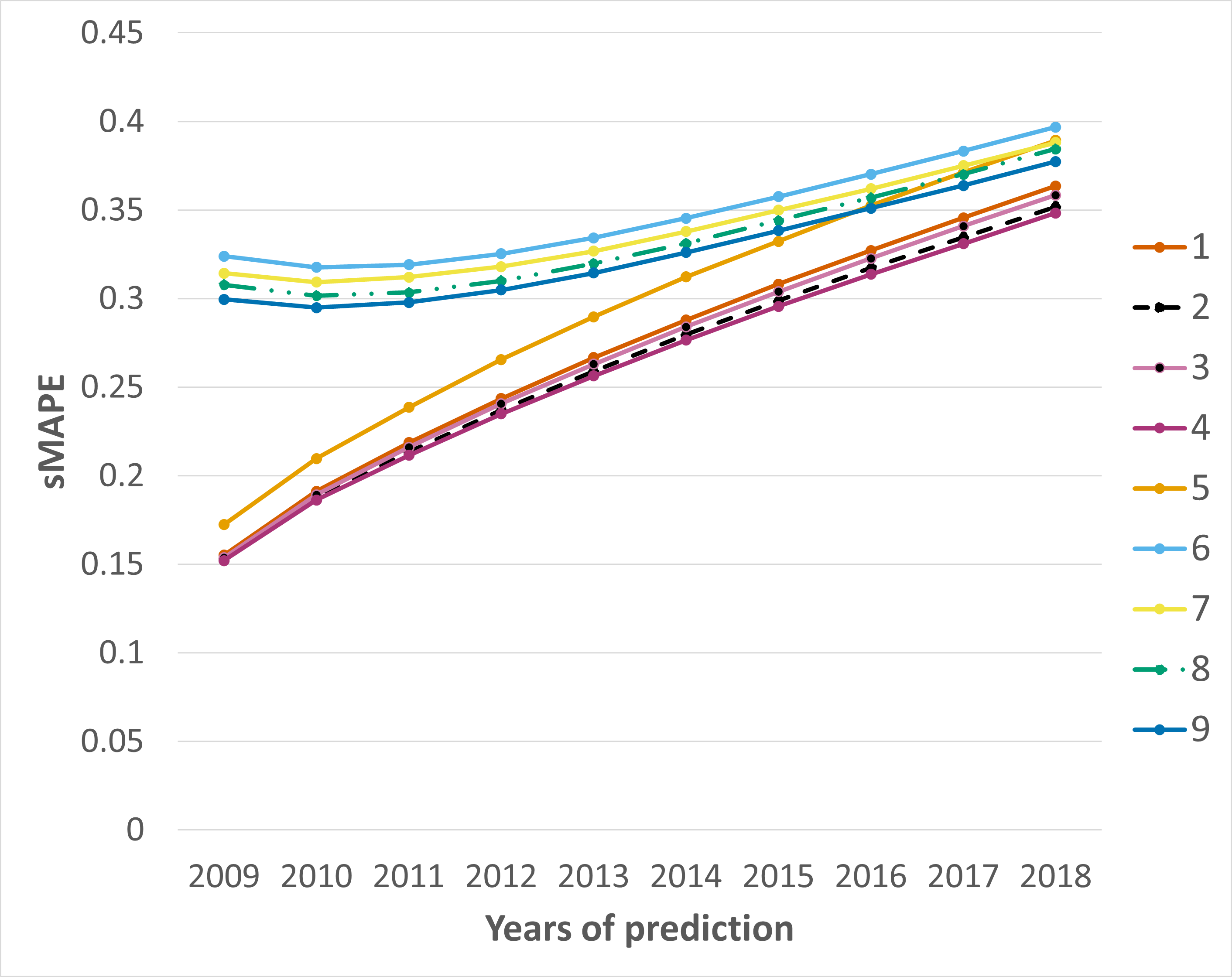}&
\includegraphics[width=.37\linewidth]{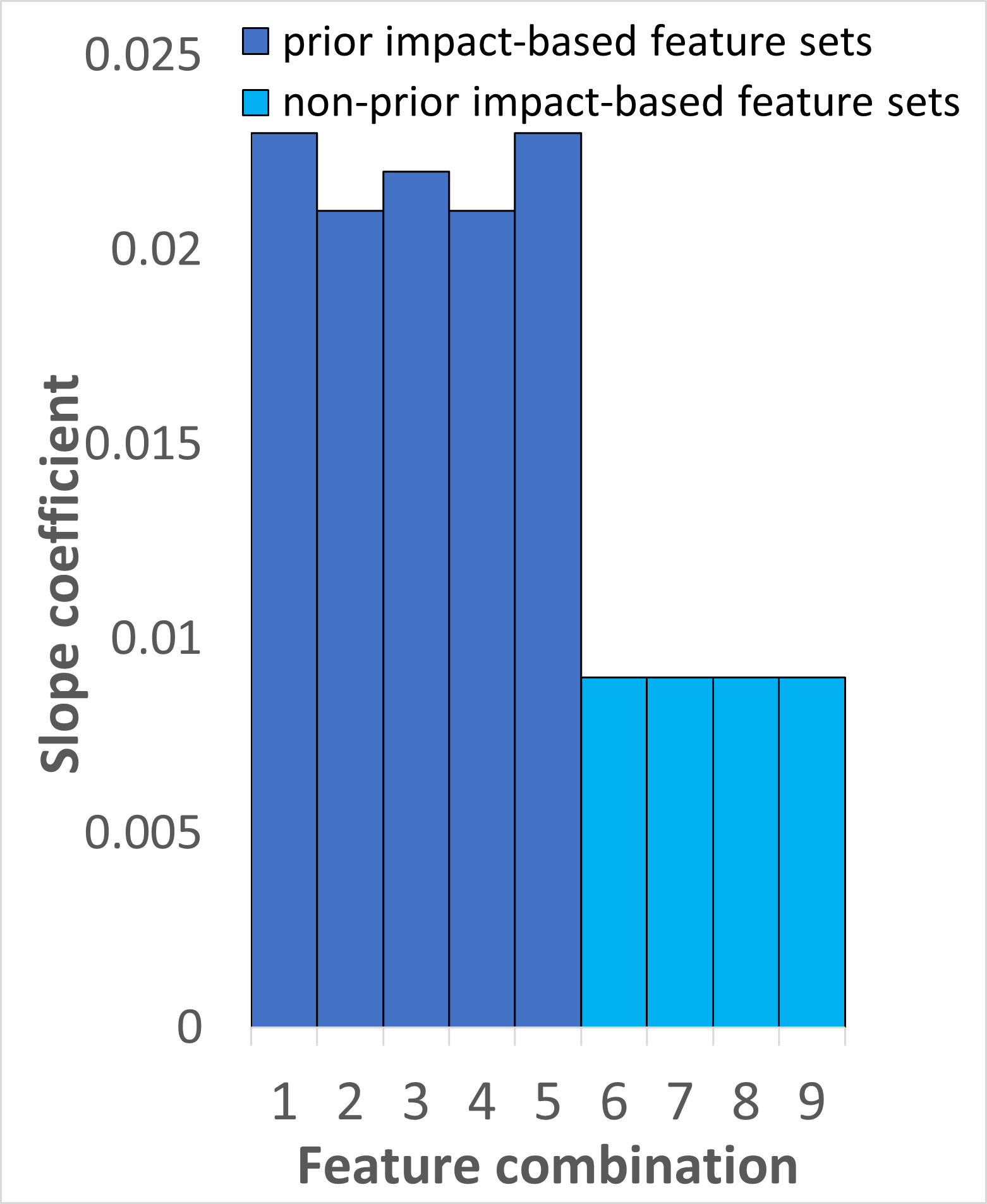}\\[-1ex]

\rowname{Senior}&
\includegraphics[width=.57\linewidth]{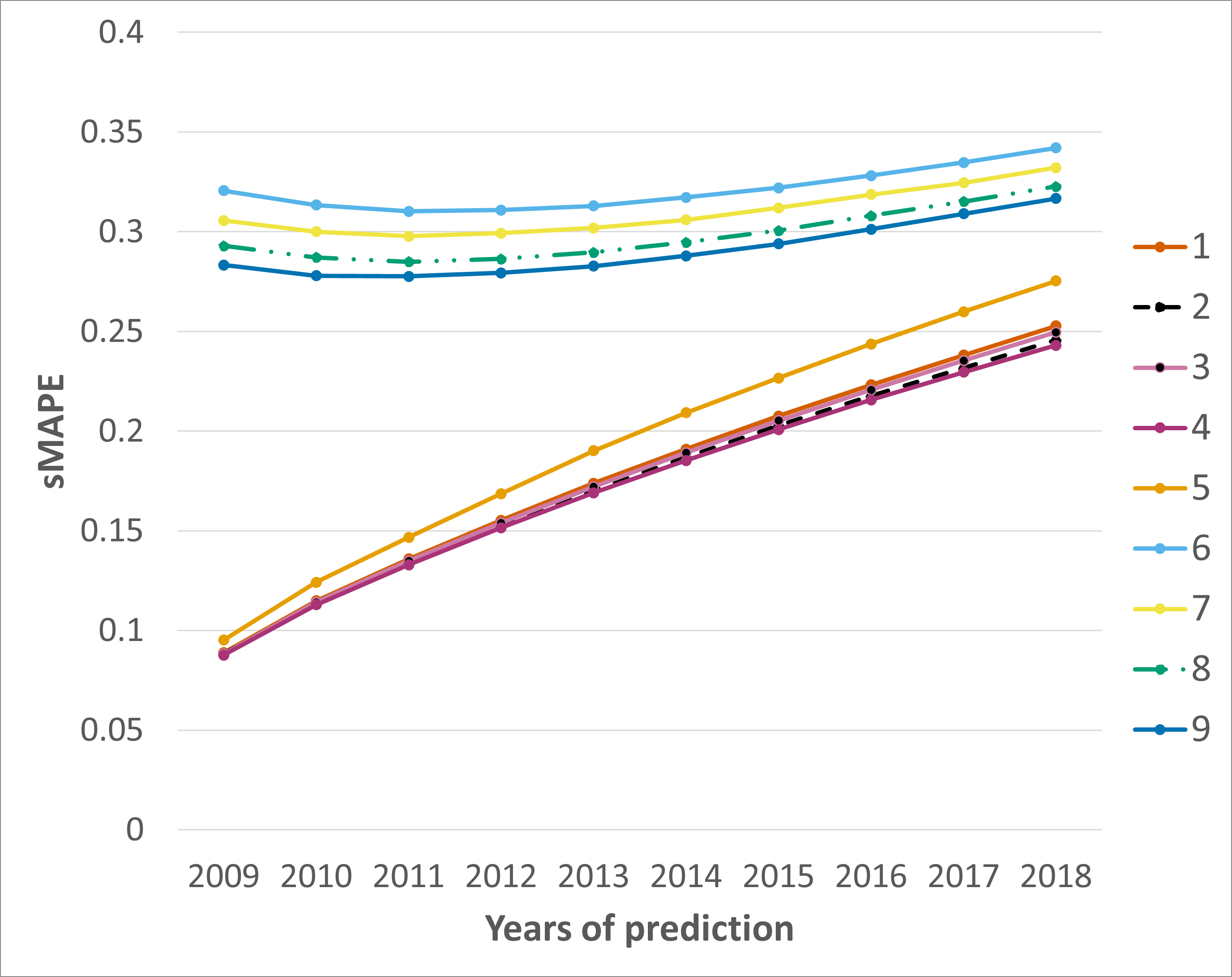}&
\includegraphics[width=.37\linewidth]{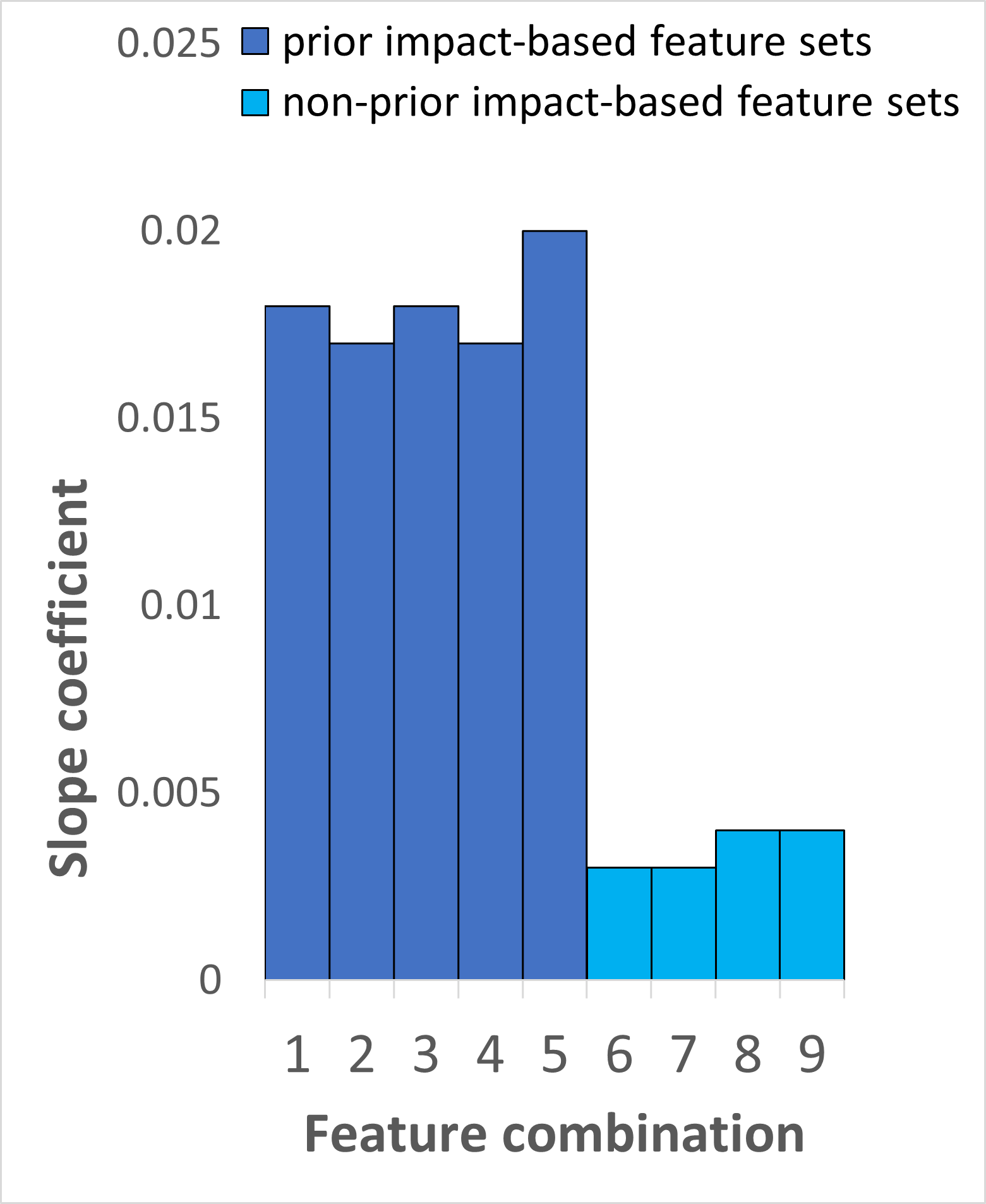}\\[-1ex]

\end{tabular}
\caption{Comparison of predictive performance (a) and slope coefficients (b) over ten years for different feature combinations trained with the XGBoost regression method among researchers of varying experience levels (junior, mid-level, and senior). (a) illustrates the performance of predicting models using the sMAPE metric. (b) displays the corresponding slope coefficients, indicating the performance change over time. The dark/light blue columns in (b) represent feature combinations, including/excluding prior impact-based features.}%
\label{figure_mape_seniorit}
\end{figure}

To further illustrate the performance trends over time, Figure \ref{figure_mape_seniorit} focuses on the sMAPE metric and covers the years from 2009 to 2018. It offers a visual representation of the prediction efficiency of different combination sets for researchers at different career stages throughout the entire time span. In this figure, the lower sMAPE for combinations including prior impact-based features indicates the higher performance for these sets, but losing the performance with the passing years for these combinations is more than other sets. 

To compare the prediction efficiency between different career stages, we implemented the prediction model for authors from three career stages and presented the performance (sMAPE) in Figure \ref{figure_mape_seniorit} (a). We observe a better performance for the combination sets containing prior impact-based features for all researchers' groups in the near future. Still, they lose more performance than sets without prior impact-based features in the distant future. Interestingly, the performance of non-prior impact-based models (e.g., sets 8 and 9) for junior researchers, which is worse than prior impact-based models (e.g., sets 1 and 5) in the earlier years, dominates them in the long term. We see a similar result for researchers at the mid-level (better performance for sets 8 and 9 than set 5). This suggests that non-prior impact-based features are more reliable in predicting the future h-index of younger researchers over distant periods.

To quantify the extent of performance degradation for the two groups of sets (prior and non-prior impact-based features), we calculated the slope coefficient for model performances reported in Figure \ref{figure_mape_seniorit} (a). The slope coefficient ($m$) was computed using the least squares method \cite{newbold2013statistics} with the following equation:

\begin{equation}
\hspace{3cm} 
\begin{gathered}
\text{$m$} = \frac{\sum(x-\bar{x}y-\bar{y})}{\sum(x-\bar{x})^2}
\end{gathered}
\label{equSlope}
\end{equation}

where $x$ represents the years from 2009 to 2018, $y$ represents the sMAPE in the corresponding year and $\bar{x}$ and $\bar{y}$ are their respective averages over the ten-year period. 

The presented slope coefficient in Figure \ref{figure_mape_seniorit} (b) reveals insights into the stability of the models' performance. A lower slope coefficient signifies greater stability, indicating that the model's performance changes more slowly and consistently over the ten-year period. Conversely, a higher slope coefficient indicates that the model's performance fluctuates more significantly.

In general, we observed a higher slope coefficient (indicating more significant performance loss over time) for feature combinations with prior impact-based features (in dark blue) compared to other feature combinations for researchers at any career stage. The lower value for sets containing non-prior impact-based features (in light blue) indicates that they are more stable predictors in the long term, although at a modest performance level.

\section{Limitations}
In this study, we considered just journal papers and not conference papers, and it causes bias issues, especially for disciplines in which authors publish their studies mainly as conference proceedings papers. Another limitation is the problem concerning data reliability and validity in calculating the features. For example, to obtain the proportion of open-access publications, we identified the access form of articles in 2019 on Unpaywall. Many journals have changed their business model to open-access or closed-access. We can not be sure about the accessibility of papers at the time of publishing and two years time windows that we considered to calculate the number of received citations. Also, we measured the mobility feature similar to our previous paper \cite{momeni2022many}, and the mentioned limitations in that paper exist for this feature too. 

\section{Conclusion and Discussion}\label{secCD}
In this study, we comprehensively investigated the impact of different feature categories on predicting the h-index for researchers at various career stages. By employing a machine learning approach and extensive feature analysis, our main objective was to understand the factors influencing researchers' future scholarly impact and how these factors differ based on their career stage.

The contributions of this research are threefold, as outlined in the introduction. Firstly, we explored the impact of various features on predicting researchers' h-index across different career stages by employing the feature selection technique, RFE, and implementing predictive models for various feature sets. This analysis gave us valuable insights into the predictive power of different attributes and their varying effectiveness at different career phases. Our analysis of Table \ref{TablePerformanceHindex} and Figure \ref{figure_mape_seniorit} (a) revealed that models with prior impact-based features demonstrated better performance than those without these features. This finding suggests that prior impact-based features are more reliable predictors of future scholarly impact, particularly for researchers in later career stages, both in the short and long term. Conversely, the smaller performance gap between models with prior impact-based feature sets and models without such features for junior researchers in the short term, and the superiority of models with non-prior impact-based features over models with prior impact-based features in the long term (as shown in Table 7), indicates that non-prior impact-based features play a more prominent role, particularly in long-term predictions, for younger researchers. This implies that these non-prior impact-based features could be valuable for identifying rising stars with strong potential for future scientific impact.

Secondly, our investigation delved into the temporal dimension of feature performance, encompassing both prior impact-based and non-prior impact-based features. We made notable observations by examining different feature combinations and their predictive power over time. Prior impact-based features exhibited the highest predictive accuracy in the short term, but their performance significantly declined in the long term compared to other features. This finding underscores the importance of considering non-prior impact-based features for enhancing long-term predictions.

Lastly, we introduced novel author (e.g., demographic characteristics) and paper/venue-specific features to estimate the author's h-index and assessed their impact on prediction tasks through feature selection analysis. The results revealed interesting insights into the individual contributions of these features to researchers' scientific impact. Among the introduced features, gender showed the weakest predictive power, suggesting that gender has almost no impact on the scientific impact, which is desirable. However, \textit{OpenAccessRatio} emerged as one of the top five powerful predictors for junior and mid-level seniors in the short term and held a similar position for seniors in the long term. In contrast, \textit{DisciplineMobility} ranked as the second top predictor for researchers from any career stage in the short term but exhibited weaker predictive power in the long term. The higher ranking of \textit{MaxCoauthorHindex} in predicting the h-index for researchers in earlier career stages, both in the short and long term, highlighted the significance of co-authors and their reputation in forecasting future h-index values. Additionally, \textit{InternationalCoauthorRatio} was among the top five predictors for mid-level researchers in the long term, while the \textit{FieldOfStudy} also held a place among the top five predictors, indicating a strong association of the h-index with specific research fields. Notably, \textit{SocialSciences} featured as one of the top predictors for senior researchers, while \textit{PhysicalSciences} played a similar role for junior and mid-level researchers in the long term, suggesting that predicting the h-index of seniors and certain disciplines in the long term is more feasible. On the other hand, \textit{MobilityScore} demonstrated no significant impact on the h-index for any of the three groups of researchers, except for mid-level researchers in the long term, where it ranked fourth. Finally, other newly introduced features, such as \textit{KeywordPopularity} and \textit{PrimaryAuthorRatio}, had minimal impact due to their low ranking in the feature selection process.

Additionally, the results of the correlation analysis were consistent with the feature selection findings. A positive moderate correlation coefficient was observed between the authors' international mobility and their future h-index. However, given the low proportion of mobile researchers (about 27\%), this author's feature proved less effective in predicting the h-index when accounting for other factors. Conversely, we found a very weak correlation between gender and the h-index, with gender displaying the lowest importance in predicting the h-index among all features. The results also underscored the importance of focusing on the study's field to achieve a better scientific impact. Paper/venue-specific features were shown to have more impact on the future h-index than the author's demographic and co-authorship characteristics.

The performances of proposed models indicate that still more features that don't depend on the history of publications and citations are required to forecast the future h-index of young researchers. For example, \cite{mccarty2013predicting, nikolentzos2021can} focused on analyzing the co-authorship network to investigate the relationship between the structural role of authors in the network and the future h-index. Using such intensive network analysis in our study could improve the performance, particularly for junior researchers with lower impact history in their profiles. Additionally, the textual content of papers examined by \cite{nikolentzos2021can} and topic authority by \cite{dong2015will} could be combined with the introduced features in this study to enhance the predictive power of our models. By incorporating these additional features alongside the ones introduced in our research, we may offer a more comprehensive understanding of researchers' future scholarly impact and lead to more accurate predictions for early-career academics.

This study aims to reveal the factors associated with the future h-index of researchers based on bibliometric data, which allowed us to have various researchers groups from different countries and scientific fields for more comprehensive analyses. The results can be informative for researchers to understand how bibliometric characteristics of authors and papers can influence the future h-index and for policymakers to support them by focusing on the factors having positive relations with scientific success.
We admit that the h-index, which is the most popular metric to assess the scholars, suffers from some limitations (e.g., field-dependent \cite{grech2018increasing}, incapable of comparing researchers in different career stages \cite{kelly2006h}  and detect authors with extremely highly cited papers  \cite{egghe2006theory}, can be manipulated by self-citations \cite{bartneck2011detecting}). Our work is not about promoting the h-index, but acknowledging its deficiencies to better understand what factors influence it. Without understanding these factors, researchers cannot understand its biases. Hence we actually contribute to understanding the deficiencies. In addition, possible bias by missing data (e.g., including only authors with gender status) can affect the validity of models. In addition, margin error has not been indicated in this study, and the reliability level of these models is uncertain.  

To predict the scientific impact, we employed artificial intelligence (AI) models, which are supposed to mimic human decision-making for assessment and don't necessarily lead to ethical and desirable results. One ethical issue is considering certain features that cause discriminatory effects or introduce bias against certain groups in the predicting model \cite{asaro2019ai,zuiderveen2018discrimination}, which we don't intend in this study. For example, investigating gender as a predictor in the prediction model was to study gender inequality in science for more attention in policy-making.

\begin{backmatter}
\section*{Declarations}
\subsection*{Competing interests}
  The authors declare that they have no competing interests.

\subsection*{Funding}
This work is financially supported by BMBF project OASE, grant number 01PU17005A. 

\subsection*{Availability of data and materials}
We don’t have permission to redistribute Spocus’s raw data, but processed data used for the analyses are available and documented in the  Git repository \href{https://github.com/momenifi/hindex/blob/main/README.md} {Git repository}.

\subsection*{Abbreviations}
\textbf{GB:} Gradient Boosting \newline
\textbf{GBRT:} Gradient Boosted Regression Trees\newline
\textbf{GBDT:} Gradient-Boosting Decision Tree\newline
\textbf{GDP:} Gross Domestic Product\newline
\textbf{KNN:} K-nearest neighbour\newline
\textbf{MAPE:} Mean Absolute Percentage Error\newline
\textbf{NN:} Neural Networks\newline
\textbf{OACA:} Open Access Citation Advantage\newline
\textbf{RF:} Random Forest \newline
\textbf{RFE:} Recursive Feature Elimination\newline
\textbf{RMSE:} Root Mean Square Error\newline
\textbf{sMAPE:} symmetric Mean Absolute Percentage Error\newline
\textbf{SVR:} Support Vector Regression\newline
\textbf{wPR:} weighted Percentile Ranking\newline
\textbf{XGBoost:} Extreme Gradient Boosting\newline

\subsection*{Author's contributions}
   Stefan Dietze supervised this study, and Philipp Mayr was the project leader that supported it financially. Material preparation, data collection, Methodology, analysis, Validation, and Visualization were performed by Fakhri Momeni. Fakhri Momeni wrote the first draft of the manuscript, and all authors commented on previous versions. All authors read and approved the final manuscript. 
   
\subsection*{Authors' information}
Fakhri Momeni is a research associate at GESIS -- Leibniz Institute for the Social Sciences in Cologne and Ph.D. student in information science at Heinrich Heine University in Duesseldorf. 

Dr. Philipp Mayr is a team leader (Information $\&$ Data Retrieval) at GESIS in Cologne, department Knowledge Technologies for the Social Sciences (KTS).

Prof. Dr. Stefan Dietze is Professor of Data $\&$ Knowledge Engineering at Heinrich Heine University Duesseldorf and Scientific Director of the Knowledge Technologies department for the Social Sciences at GESIS in Cologne.

\subsection*{Acknowledgements}
We acknowledge
the support of the German Competence Center for Bibliometrics (grant: 01PQ17001) for maintaining the used dataset for the analyses.


\bibliographystyle{bmc-mathphys} 
\bibliography{bmc_article}      


\begin{thebibliography}{57}
\ifx \bisbn   \undefined \def \bisbn  #1{ISBN #1}\fi
\ifx \binits  \undefined \def \binits#1{#1}\fi
\ifx \bauthor  \undefined \def \bauthor#1{#1}\fi
\ifx \batitle  \undefined \def \batitle#1{#1}\fi
\ifx \bjtitle  \undefined \def \bjtitle#1{#1}\fi
\ifx \bvolume  \undefined \def \bvolume#1{\textbf{#1}}\fi
\ifx \byear  \undefined \def \byear#1{#1}\fi
\ifx \bissue  \undefined \def \bissue#1{#1}\fi
\ifx \bfpage  \undefined \def \bfpage#1{#1}\fi
\ifx \blpage  \undefined \def \blpage #1{#1}\fi
\ifx \burl  \undefined \def \burl#1{\textsf{#1}}\fi
\ifx \doiurl  \undefined \def \doiurl#1{\textsf{#1}}\fi
\ifx \betal  \undefined \def \betal{\textit{et al.}}\fi
\ifx \binstitute  \undefined \def \binstitute#1{#1}\fi
\ifx \binstitutionaled  \undefined \def \binstitutionaled#1{#1}\fi
\ifx \bctitle  \undefined \def \bctitle#1{#1}\fi
\ifx \beditor  \undefined \def \beditor#1{#1}\fi
\ifx \bpublisher  \undefined \def \bpublisher#1{#1}\fi
\ifx \bbtitle  \undefined \def \bbtitle#1{#1}\fi
\ifx \bedition  \undefined \def \bedition#1{#1}\fi
\ifx \bseriesno  \undefined \def \bseriesno#1{#1}\fi
\ifx \blocation  \undefined \def \blocation#1{#1}\fi
\ifx \bsertitle  \undefined \def \bsertitle#1{#1}\fi
\ifx \bsnm \undefined \def \bsnm#1{#1}\fi
\ifx \bsuffix \undefined \def \bsuffix#1{#1}\fi
\ifx \bparticle \undefined \def \bparticle#1{#1}\fi
\ifx \barticle \undefined \def \barticle#1{#1}\fi
\ifx \bconfdate \undefined \def \bconfdate #1{#1}\fi
\ifx \botherref \undefined \def \botherref #1{#1}\fi
\ifx \url \undefined \def \url#1{\textsf{#1}}\fi
\ifx \bchapter \undefined \def \bchapter#1{#1}\fi
\ifx \bbook \undefined \def \bbook#1{#1}\fi
\ifx \bcomment \undefined \def \bcomment#1{#1}\fi
\ifx \oauthor \undefined \def \oauthor#1{#1}\fi
\ifx \citeauthoryear \undefined \def \citeauthoryear#1{#1}\fi
\ifx \endbibitem  \undefined \def \endbibitem {}\fi
\ifx \bconflocation  \undefined \def \bconflocation#1{#1}\fi
\ifx \arxivurl  \undefined \def \arxivurl#1{\textsf{#1}}\fi
\csname PreBibitemsHook\endcsname

\bibitem{hirsch2005index}
\begin{barticle}
\bauthor{\bsnm{Hirsch}, \binits{J.E.}}:
\batitle{An index to quantify an individual's scientific research output}.
\bjtitle{Proceedings of the National academy of Sciences}
\bvolume{102}(\bissue{46}),
\bfpage{16569}--\blpage{16572}
(\byear{2005})
\end{barticle}
\endbibitem

\bibitem{egghe2006improvement}
\begin{barticle}
\bauthor{\bsnm{Egghe}, \binits{L.}}, \betal:
\batitle{An improvement of the h-index: The g-index}.
\bjtitle{ISSI newsletter}
\bvolume{2}(\bissue{1}),
\bfpage{8}--\blpage{9}
(\byear{2006})
\end{barticle}
\endbibitem

\bibitem{kaur2013universality}
\begin{barticle}
\bauthor{\bsnm{Kaur}, \binits{J.}},
\bauthor{\bsnm{Radicchi}, \binits{F.}},
\bauthor{\bsnm{Menczer}, \binits{F.}}:
\batitle{Universality of scholarly impact metrics}.
\bjtitle{Journal of Informetrics}
\bvolume{7}(\bissue{4}),
\bfpage{924}--\blpage{932}
(\byear{2013})
\end{barticle}
\endbibitem

\bibitem{daud2013finding}
\begin{bchapter}
\bauthor{\bsnm{Daud}, \binits{A.}},
\bauthor{\bsnm{Abbasi}, \binits{R.}},
\bauthor{\bsnm{Muhammad}, \binits{F.}}:
\bctitle{Finding rising stars in social networks}.
In: \bbtitle{International Conference on Database Systems for Advanced
  Applications},
pp. \bfpage{13}--\blpage{24}
(\byear{2013}).
\bcomment{Springer}
\end{bchapter}
\endbibitem

\bibitem{ayaz2018predicting}
\begin{barticle}
\bauthor{\bsnm{Ayaz}, \binits{S.}},
\bauthor{\bsnm{Masood}, \binits{N.}},
\bauthor{\bsnm{Islam}, \binits{M.A.}}:
\batitle{Predicting scientific impact based on h-index}.
\bjtitle{Scientometrics}
\bvolume{114}(\bissue{3}),
\bfpage{993}--\blpage{1010}
(\byear{2018})
\end{barticle}
\endbibitem

\bibitem{weihs2017learning}
\begin{bchapter}
\bauthor{\bsnm{Weihs}, \binits{L.}},
\bauthor{\bsnm{Etzioni}, \binits{O.}}:
\bctitle{Learning to predict citation-based impact measures}.
In: \bbtitle{2017 ACM/IEEE Joint Conference on Digital Libraries (JCDL)},
pp. \bfpage{1}--\blpage{10}
(\byear{2017}).
\bcomment{IEEE}
\end{bchapter}
\endbibitem

\bibitem{wu2019predicting}
\begin{barticle}
\bauthor{\bsnm{Wu}, \binits{Z.}},
\bauthor{\bsnm{Lin}, \binits{W.}},
\bauthor{\bsnm{Liu}, \binits{P.}},
\bauthor{\bsnm{Chen}, \binits{J.}},
\bauthor{\bsnm{Mao}, \binits{L.}}:
\batitle{Predicting long-term scientific impact based on multi-field feature
  extraction}.
\bjtitle{IEEE Access}
\bvolume{7},
\bfpage{51759}--\blpage{51770}
(\byear{2019})
\end{barticle}
\endbibitem

\bibitem{bai2019predicting}
\begin{barticle}
\bauthor{\bsnm{Bai}, \binits{X.}},
\bauthor{\bsnm{Zhang}, \binits{F.}},
\bauthor{\bsnm{Lee}, \binits{I.}}:
\batitle{Predicting the citations of scholarly paper}.
\bjtitle{Journal of Informetrics}
\bvolume{13}(\bissue{1}),
\bfpage{407}--\blpage{418}
(\byear{2019})
\end{barticle}
\endbibitem

\bibitem{abrishami2019predicting}
\begin{barticle}
\bauthor{\bsnm{Abrishami}, \binits{A.}},
\bauthor{\bsnm{Aliakbary}, \binits{S.}}:
\batitle{Predicting citation counts based on deep neural network learning
  techniques}.
\bjtitle{Journal of Informetrics}
\bvolume{13}(\bissue{2}),
\bfpage{485}--\blpage{499}
(\byear{2019})
\end{barticle}
\endbibitem

\bibitem{jiang2021hints}
\begin{bchapter}
\bauthor{\bsnm{Jiang}, \binits{S.}},
\bauthor{\bsnm{Koch}, \binits{B.}},
\bauthor{\bsnm{Sun}, \binits{Y.}}:
\bctitle{Hints: Citation time series prediction for new publications via
  dynamic heterogeneous information network embedding}.
In: \bbtitle{Proceedings of the Web Conference 2021},
pp. \bfpage{3158}--\blpage{3167}
(\byear{2021})
\end{bchapter}
\endbibitem

\bibitem{ruan2020predicting}
\begin{barticle}
\bauthor{\bsnm{Ruan}, \binits{X.}},
\bauthor{\bsnm{Zhu}, \binits{Y.}},
\bauthor{\bsnm{Li}, \binits{J.}},
\bauthor{\bsnm{Cheng}, \binits{Y.}}:
\batitle{Predicting the citation counts of individual papers via a bp neural
  network}.
\bjtitle{Journal of Informetrics}
\bvolume{14}(\bissue{3}),
\bfpage{101039}
(\byear{2020})
\end{barticle}
\endbibitem

\bibitem{kossmeier2019predicting}
\begin{barticle}
\bauthor{\bsnm{Kossmeier}, \binits{M.}},
\bauthor{\bsnm{Heinze}, \binits{G.}}:
\batitle{Predicting future citation counts of scientific manuscripts submitted
  for publication: a cohort study in transplantology}.
\bjtitle{Transplant International}
\bvolume{32}(\bissue{1}),
\bfpage{6}--\blpage{15}
(\byear{2019})
\end{barticle}
\endbibitem

\bibitem{nikolentzos2021can}
\begin{botherref}
\oauthor{\bsnm{Nikolentzos}, \binits{G.}},
\oauthor{\bsnm{Panagopoulos}, \binits{G.}},
\oauthor{\bsnm{Evdaimon}, \binits{I.}},
\oauthor{\bsnm{Vazirgiannis}, \binits{M.}}:
Can author collaboration reveal impact? the case of h-index,
177--194
(2021)
\end{botherref}
\endbibitem

\bibitem{nie2019academic}
\begin{barticle}
\bauthor{\bsnm{Nie}, \binits{Y.}},
\bauthor{\bsnm{Zhu}, \binits{Y.}},
\bauthor{\bsnm{Lin}, \binits{Q.}},
\bauthor{\bsnm{Zhang}, \binits{S.}},
\bauthor{\bsnm{Shi}, \binits{P.}},
\bauthor{\bsnm{Niu}, \binits{Z.}}:
\batitle{Academic rising star prediction via scholar’s evaluation model and
  machine learning techniques}.
\bjtitle{Scientometrics}
\bvolume{120}(\bissue{2}),
\bfpage{461}--\blpage{476}
(\byear{2019})
\end{barticle}
\endbibitem

\bibitem{mccarty2013predicting}
\begin{barticle}
\bauthor{\bsnm{McCarty}, \binits{C.}},
\bauthor{\bsnm{Jawitz}, \binits{J.W.}},
\bauthor{\bsnm{Hopkins}, \binits{A.}},
\bauthor{\bsnm{Goldman}, \binits{A.}}:
\batitle{Predicting author h-index using characteristics of the co-author
  network}.
\bjtitle{Scientometrics}
\bvolume{96}(\bissue{2}),
\bfpage{467}--\blpage{483}
(\byear{2013})
\end{barticle}
\endbibitem

\bibitem{dong2016can}
\begin{barticle}
\bauthor{\bsnm{Dong}, \binits{Y.}},
\bauthor{\bsnm{Johnson}, \binits{R.A.}},
\bauthor{\bsnm{Chawla}, \binits{N.V.}}:
\batitle{Can scientific impact be predicted?}
\bjtitle{IEEE Transactions on Big Data}
\bvolume{2}(\bissue{1}),
\bfpage{18}--\blpage{30}
(\byear{2016})
\end{barticle}
\endbibitem

\bibitem{momeni2022many}
\begin{barticle}
\bauthor{\bsnm{Momeni}, \binits{F.}},
\bauthor{\bsnm{Karimi}, \binits{F.}},
\bauthor{\bsnm{Mayr}, \binits{P.}},
\bauthor{\bsnm{Peters}, \binits{I.}},
\bauthor{\bsnm{Dietze}, \binits{S.}}:
\batitle{The many facets of academic mobility and its impact on scholars'
  career}.
\bjtitle{Journal of Informetrics}
\bvolume{16}(\bissue{2}),
\bfpage{101280}
(\byear{2022})
\end{barticle}
\endbibitem

\bibitem{singh2018comparing}
\begin{barticle}
\bauthor{\bsnm{Singh}, \binits{V.}}:
\batitle{Comparing research productivity of returnee-phds in science,
  engineering, and the social sciences}.
\bjtitle{Scientometrics}
\bvolume{115}(\bissue{3}),
\bfpage{1241}--\blpage{1252}
(\byear{2018})
\end{barticle}
\endbibitem

\bibitem{netz2020effects}
\begin{barticle}
\bauthor{\bsnm{Netz}, \binits{N.}},
\bauthor{\bsnm{Hampel}, \binits{S.}},
\bauthor{\bsnm{Aman}, \binits{V.}}:
\batitle{What effects does international mobility have on scientists’
  careers? a systematic review}.
\bjtitle{Research evaluation}
\bvolume{29}(\bissue{3}),
\bfpage{327}--\blpage{351}
(\byear{2020})
\end{barticle}
\endbibitem

\bibitem{liu2021academic}
\begin{barticle}
\bauthor{\bsnm{Liu}, \binits{J.}},
\bauthor{\bsnm{Wang}, \binits{R.}},
\bauthor{\bsnm{Xu}, \binits{S.}}:
\batitle{What academic mobility configurations contribute to high performance:
  an fsqca analysis of csc-funded visiting scholars}.
\bjtitle{Scientometrics}
\bvolume{126}(\bissue{2}),
\bfpage{1079}--\blpage{1100}
(\byear{2021})
\end{barticle}
\endbibitem

\bibitem{radford2022h}
\begin{barticle}
\bauthor{\bsnm{Radford}, \binits{D.M.}},
\bauthor{\bsnm{Parangi}, \binits{S.}},
\bauthor{\bsnm{Tu}, \binits{C.}},
\bauthor{\bsnm{Silver}, \binits{J.K.}}:
\batitle{h-index and academic rank by gender among breast surgery fellowship
  faculty}.
\bjtitle{Journal of Women's Health}
\bvolume{31}(\bissue{1}),
\bfpage{110}--\blpage{116}
(\byear{2022})
\end{barticle}
\endbibitem

\bibitem{carter2017gender}
\begin{barticle}
\bauthor{\bsnm{Carter}, \binits{T.E.}},
\bauthor{\bsnm{Smith}, \binits{T.E.}},
\bauthor{\bsnm{Osteen}, \binits{P.J.}}:
\batitle{Gender comparisons of social work faculty using h-index scores}.
\bjtitle{Scientometrics}
\bvolume{111}(\bissue{3}),
\bfpage{1547}--\blpage{1557}
(\byear{2017})
\end{barticle}
\endbibitem

\bibitem{lopez2014gender}
\begin{barticle}
\bauthor{\bsnm{Lopez}, \binits{S.A.}},
\bauthor{\bsnm{Svider}, \binits{P.F.}},
\bauthor{\bsnm{Misra}, \binits{P.}},
\bauthor{\bsnm{Bhagat}, \binits{N.}},
\bauthor{\bsnm{Langer}, \binits{P.D.}},
\bauthor{\bsnm{Eloy}, \binits{J.A.}}:
\batitle{Gender differences in promotion and scholarly impact: an analysis of
  1460 academic ophthalmologists}.
\bjtitle{Journal of surgical education}
\bvolume{71}(\bissue{6}),
\bfpage{851}--\blpage{859}
(\byear{2014})
\end{barticle}
\endbibitem

\bibitem{kelly2006h}
\begin{barticle}
\bauthor{\bsnm{Kelly}, \binits{C.D.}},
\bauthor{\bsnm{Jennions}, \binits{M.D.}}:
\batitle{The h index and career assessment by numbers}.
\bjtitle{Trends in Ecology \& Evolution}
\bvolume{21}(\bissue{4}),
\bfpage{167}--\blpage{170}
(\byear{2006})
\end{barticle}
\endbibitem

\bibitem{leydesdorff2019relative}
\begin{barticle}
\bauthor{\bsnm{Leydesdorff}, \binits{L.}},
\bauthor{\bsnm{Bornmann}, \binits{L.}},
\bauthor{\bsnm{Wagner}, \binits{C.S.}}:
\batitle{The relative influences of government funding and international
  collaboration on citation impact}.
\bjtitle{Journal of the Association for Information Science and Technology}
\bvolume{70}(\bissue{2}),
\bfpage{198}--\blpage{201}
(\byear{2019})
\end{barticle}
\endbibitem

\bibitem{smirnova2023comprehensive}
\begin{barticle}
\bauthor{\bsnm{Smirnova}, \binits{N.}},
\bauthor{\bsnm{Mayr}, \binits{P.}}:
\batitle{A comprehensive analysis of acknowledgement texts in web of science: a
  case study on four scientific domains}.
\bjtitle{Scientometrics}
\bvolume{128}(\bissue{1}),
\bfpage{709}--\blpage{734}
(\byear{2023})
\end{barticle}
\endbibitem

\bibitem{gantman2012economic}
\begin{barticle}
\bauthor{\bsnm{Gantman}, \binits{E.R.}}:
\batitle{Economic, linguistic, and political factors in the scientific
  productivity of countries}.
\bjtitle{Scientometrics}
\bvolume{93}(\bissue{3}),
\bfpage{967}--\blpage{985}
(\byear{2012})
\end{barticle}
\endbibitem

\bibitem{confraria2017determinants}
\begin{barticle}
\bauthor{\bsnm{Confraria}, \binits{H.}},
\bauthor{\bsnm{Godinho}, \binits{M.M.}},
\bauthor{\bsnm{Wang}, \binits{L.}}:
\batitle{Determinants of citation impact: A comparative analysis of the global
  south versus the global north}.
\bjtitle{Research Policy}
\bvolume{46}(\bissue{1}),
\bfpage{265}--\blpage{279}
(\byear{2017})
\end{barticle}
\endbibitem

\bibitem{malesios2014comparison}
\begin{barticle}
\bauthor{\bsnm{Malesios}, \binits{C.}},
\bauthor{\bsnm{Psarakis}, \binits{S.}}:
\batitle{Comparison of the h-index for different fields of research using
  bootstrap methodology}.
\bjtitle{Quality \& Quantity}
\bvolume{48}(\bissue{1}),
\bfpage{521}--\blpage{545}
(\byear{2014})
\end{barticle}
\endbibitem

\bibitem{lillquist2010discipline}
\begin{barticle}
\bauthor{\bsnm{Lillquist}, \binits{E.}},
\bauthor{\bsnm{Green}, \binits{S.}}:
\batitle{The discipline dependence of citation statistics}.
\bjtitle{Scientometrics}
\bvolume{84}(\bissue{3}),
\bfpage{749}--\blpage{762}
(\byear{2010})
\end{barticle}
\endbibitem

\bibitem{iglesias2007scaling}
\begin{barticle}
\bauthor{\bsnm{Iglesias}, \binits{J.}},
\bauthor{\bsnm{Pecharrom{\'a}n}, \binits{C.}}:
\batitle{Scaling the h-index for different scientific isi fields}.
\bjtitle{Scientometrics}
\bvolume{73}(\bissue{3}),
\bfpage{303}--\blpage{320}
(\byear{2007})
\end{barticle}
\endbibitem

\bibitem{petersen2014inequality}
\begin{barticle}
\bauthor{\bsnm{Petersen}, \binits{A.M.}},
\bauthor{\bsnm{Penner}, \binits{O.}}:
\batitle{Inequality and cumulative advantage in science careers: a case study
  of high-impact journals}.
\bjtitle{EPJ Data Science}
\bvolume{3},
\bfpage{1}--\blpage{25}
(\byear{2014})
\end{barticle}
\endbibitem

\bibitem{xie2022open}
\begin{barticle}
\bauthor{\bsnm{Xie}, \binits{F.}},
\bauthor{\bsnm{Ghozy}, \binits{S.}},
\bauthor{\bsnm{Kallmes}, \binits{D.F.}},
\bauthor{\bsnm{Lehman}, \binits{J.S.}}:
\batitle{Do open-access dermatology articles have higher citation counts than
  those with subscription-based access?}
\bjtitle{PloS one}
\bvolume{17}(\bissue{12}),
\bfpage{0279265}
(\byear{2022})
\end{barticle}
\endbibitem

\bibitem{blair2020open}
\begin{botherref}
\oauthor{\bsnm{Blair}, \binits{L.D.}},
\oauthor{\bsnm{Odell}, \binits{J.D.}}:
The open access policy citation advantage for a medical school
(2020)
\end{botherref}
\endbibitem

\bibitem{ottaviani2016post}
\begin{barticle}
\bauthor{\bsnm{Ottaviani}, \binits{J.}}:
\batitle{The post-embargo open access citation advantage: it exists (probably),
  it’s modest (usually), and the rich get richer (of course)}.
\bjtitle{PloS one}
\bvolume{11}(\bissue{8}),
\bfpage{0159614}
(\byear{2016})
\end{barticle}
\endbibitem

\bibitem{amjad2022investigating}
\begin{barticle}
\bauthor{\bsnm{Amjad}, \binits{T.}},
\bauthor{\bsnm{Sabir}, \binits{M.}},
\bauthor{\bsnm{Shamim}, \binits{A.}},
\bauthor{\bsnm{Amjad}, \binits{M.}},
\bauthor{\bsnm{Daud}, \binits{A.}}:
\batitle{Investigating the citation advantage of author-pays charges model in
  computer science research: a case study of elsevier and springer}.
\bjtitle{Library Hi Tech}
\bvolume{40}(\bissue{3}),
\bfpage{685}--\blpage{703}
(\byear{2022})
\end{barticle}
\endbibitem

\bibitem{langham2021open}
\begin{barticle}
\bauthor{\bsnm{Langham-Putrow}, \binits{A.}},
\bauthor{\bsnm{Bakker}, \binits{C.}},
\bauthor{\bsnm{Riegelman}, \binits{A.}}:
\batitle{Is the open access citation advantage real? a systematic review of the
  citation of open access and subscription-based articles}.
\bjtitle{PloS one}
\bvolume{16}(\bissue{6}),
\bfpage{0253129}
(\byear{2021})
\end{barticle}
\endbibitem

\bibitem{fraser2020relationship}
\begin{barticle}
\bauthor{\bsnm{Fraser}, \binits{N.}},
\bauthor{\bsnm{Momeni}, \binits{F.}},
\bauthor{\bsnm{Mayr}, \binits{P.}},
\bauthor{\bsnm{Peters}, \binits{I.}}:
\batitle{The relationship between biorxiv preprints, citations and altmetrics}.
\bjtitle{Quantitative Science Studies}
\bvolume{1}(\bissue{2}),
\bfpage{618}--\blpage{638}
(\byear{2020})
\end{barticle}
\endbibitem

\bibitem{momeni2022factors}
\begin{botherref}
\oauthor{\bsnm{Momeni}, \binits{F.}},
\oauthor{\bsnm{Dietze}, \binits{S.}},
\oauthor{\bsnm{Mayr}, \binits{P.}},
\oauthor{\bsnm{Biesenbender}, \binits{K.}},
\oauthor{\bsnm{Peters}, \binits{I.}}:
Which factors drive open access publishing? a springer nature case study.
arXiv preprint arXiv:2208.08221
(2022)
\end{botherref}
\endbibitem

\bibitem{hsu2011correlation}
\begin{barticle}
\bauthor{\bsnm{Hsu}, \binits{J.-w.}},
\bauthor{\bsnm{Huang}, \binits{D.-w.}}:
\batitle{Correlation between impact and collaboration}.
\bjtitle{Scientometrics}
\bvolume{86}(\bissue{2}),
\bfpage{317}--\blpage{324}
(\byear{2011})
\end{barticle}
\endbibitem

\bibitem{puuska2014international}
\begin{barticle}
\bauthor{\bsnm{Puuska}, \binits{H.-M.}},
\bauthor{\bsnm{Muhonen}, \binits{R.}},
\bauthor{\bsnm{Leino}, \binits{Y.}}:
\batitle{International and domestic co-publishing and their citation impact in
  different disciplines}.
\bjtitle{Scientometrics}
\bvolume{98}(\bissue{2}),
\bfpage{823}--\blpage{839}
(\byear{2014})
\end{barticle}
\endbibitem

\bibitem{sarigol2014predicting}
\begin{barticle}
\bauthor{\bsnm{Sarig{\"o}l}, \binits{E.}},
\bauthor{\bsnm{Pfitzner}, \binits{R.}},
\bauthor{\bsnm{Scholtes}, \binits{I.}},
\bauthor{\bsnm{Garas}, \binits{A.}},
\bauthor{\bsnm{Schweitzer}, \binits{F.}}:
\batitle{Predicting scientific success based on coauthorship networks}.
\bjtitle{EPJ Data Science}
\bvolume{3},
\bfpage{1}--\blpage{16}
(\byear{2014})
\end{barticle}
\endbibitem

\bibitem{ni2018relationship}
\begin{barticle}
\bauthor{\bsnm{Ni}, \binits{P.}},
\bauthor{\bsnm{An}, \binits{X.}}:
\batitle{Relationship between international collaboration papers and their
  citations from an economic perspective}.
\bjtitle{Scientometrics}
\bvolume{116}(\bissue{2}),
\bfpage{863}--\blpage{877}
(\byear{2018})
\end{barticle}
\endbibitem

\bibitem{karimi2016inferring}
\begin{bchapter}
\bauthor{\bsnm{Karimi}, \binits{F.}},
\bauthor{\bsnm{Wagner}, \binits{C.}},
\bauthor{\bsnm{Lemmerich}, \binits{F.}},
\bauthor{\bsnm{Jadidi}, \binits{M.}},
\bauthor{\bsnm{Strohmaier}, \binits{M.}}:
\bctitle{Inferring gender from names on the web: A comparative evaluation of
  gender detection methods}.
In: \bbtitle{Proceedings of the 25th International Conference Companion on
  World Wide Web},
pp. \bfpage{53}--\blpage{54}
(\byear{2016})
\end{bchapter}
\endbibitem

\bibitem{bornmann2014p100}
\begin{barticle}
\bauthor{\bsnm{Bornmann}, \binits{L.}},
\bauthor{\bsnm{Mutz}, \binits{R.}}:
\batitle{From p100 to p100': A new citation-rank approach}.
\bjtitle{Journal of the Association for Information Science and Technology}
\bvolume{65}(\bissue{9}),
\bfpage{1939}--\blpage{1943}
(\byear{2014})
\end{barticle}
\endbibitem

\bibitem{bornmann2020evaluation}
\begin{barticle}
\bauthor{\bsnm{Bornmann}, \binits{L.}},
\bauthor{\bsnm{Williams}, \binits{R.}}:
\batitle{An evaluation of percentile measures of citation impact, and a
  proposal for making them better}.
\bjtitle{Scientometrics}
\bvolume{124}(\bissue{2}),
\bfpage{1457}--\blpage{1478}
(\byear{2020})
\end{barticle}
\endbibitem

\bibitem{chen2016xgboost}
\begin{bchapter}
\bauthor{\bsnm{Chen}, \binits{T.}},
\bauthor{\bsnm{Guestrin}, \binits{C.}}:
\bctitle{Xgboost: A scalable tree boosting system}.
In: \bbtitle{Proceedings of the 22nd Acm Sigkdd International Conference on
  Knowledge Discovery and Data Mining},
pp. \bfpage{785}--\blpage{794}
(\byear{2016})
\end{bchapter}
\endbibitem

\bibitem{blasco2013using}
\begin{barticle}
\bauthor{\bsnm{Blasco}, \binits{B.C.}},
\bauthor{\bsnm{Moreno}, \binits{J.J.M.}},
\bauthor{\bsnm{Pol}, \binits{A.P.}},
\bauthor{\bsnm{Abad}, \binits{A.S.}}:
\batitle{Using the r-mape index as a resistant measure of forecast accuracy}.
\bjtitle{Psicothema}
\bvolume{25}(\bissue{4}),
\bfpage{500}--\blpage{506}
(\byear{2013})
\end{barticle}
\endbibitem

\bibitem{dong2015will}
\begin{bchapter}
\bauthor{\bsnm{Dong}, \binits{Y.}},
\bauthor{\bsnm{Johnson}, \binits{R.A.}},
\bauthor{\bsnm{Chawla}, \binits{N.V.}}:
\bctitle{Will this paper increase your h-index? scientific impact prediction}.
In: \bbtitle{Proceedings of the Eighth ACM International Conference on Web
  Search and Data Mining},
pp. \bfpage{149}--\blpage{158}
(\byear{2015})
\end{bchapter}
\endbibitem

\bibitem{artur2021review}
\begin{barticle}
\bauthor{\bsnm{Artur}, \binits{M.}}:
\batitle{Review the performance of the bernoulli na{\"\i}ve bayes classifier in
  intrusion detection systems using recursive feature elimination with
  cross-validated selection of the best number of features}.
\bjtitle{Procedia Computer Science}
\bvolume{190},
\bfpage{564}--\blpage{570}
(\byear{2021})
\end{barticle}
\endbibitem

\bibitem{zhao2022rfe}
\begin{bchapter}
\bauthor{\bsnm{Zhao}, \binits{L.}},
\bauthor{\bsnm{Deng}, \binits{F.}},
\bauthor{\bsnm{Zhang}, \binits{X.}},
\bauthor{\bsnm{Yu}, \binits{N.}}:
\bctitle{Rfe based feature selection improves performance of classifying
  multiple-causes deaths in colorectal cancer}.
In: \bbtitle{2022 7th International Conference on Intelligent Informatics and
  Biomedical Science (ICIIBMS)},
vol. \bseriesno{7},
pp. \bfpage{188}--\blpage{194}
(\byear{2022}).
\bcomment{IEEE}
\end{bchapter}
\endbibitem

\bibitem{newbold2013statistics}
\begin{botherref}
\oauthor{\bsnm{Newbold}, \binits{P.}},
\oauthor{\bsnm{Carlson}, \binits{W.L.}},
\oauthor{\bsnm{Thorne}, \binits{B.}}:
Statistics for business and economics.
Pearson
(2013)
\end{botherref}
\endbibitem

\bibitem{grech2018increasing}
\begin{botherref}
\oauthor{\bsnm{Grech}, \binits{V.}},
\oauthor{\bsnm{Rizk}, \binits{D.E.}}:
Increasing importance of research metrics: Journal Impact Factor and h-index.
Springer
(2018)
\end{botherref}
\endbibitem

\bibitem{egghe2006theory}
\begin{barticle}
\bauthor{\bsnm{Egghe}, \binits{L.}}:
\batitle{Theory and practise of the g-index}.
\bjtitle{Scientometrics}
\bvolume{69}(\bissue{1}),
\bfpage{131}--\blpage{152}
(\byear{2006})
\end{barticle}
\endbibitem

\bibitem{bartneck2011detecting}
\begin{barticle}
\bauthor{\bsnm{Bartneck}, \binits{C.}},
\bauthor{\bsnm{Kokkelmans}, \binits{S.}}:
\batitle{Detecting h-index manipulation through self-citation analysis}.
\bjtitle{Scientometrics}
\bvolume{87}(\bissue{1}),
\bfpage{85}--\blpage{98}
(\byear{2011})
\end{barticle}
\endbibitem

\bibitem{asaro2019ai}
\begin{barticle}
\bauthor{\bsnm{Asaro}, \binits{P.M.}}:
\batitle{Ai ethics in predictive policing: From models of threat to an ethics
  of care}.
\bjtitle{IEEE Technology and Society Magazine}
\bvolume{38}(\bissue{2}),
\bfpage{40}--\blpage{53}
(\byear{2019})
\end{barticle}
\endbibitem

\bibitem{zuiderveen2018discrimination}
\begin{botherref}
\oauthor{\bsnm{Zuiderveen~Borgesius}, \binits{F.}}, et al.:
Discrimination, artificial intelligence, and algorithmic decision-making.
l{\'\i}nea], Council of Europe
(2018)
\end{botherref}
\endbibitem

\end{thebibliography}

\newcommand{\BMCxmlcomment}[1]{}

\BMCxmlcomment{

<refgrp>

<bibl id="B1">
  <title><p>An index to quantify an individual's scientific research
  output</p></title>
  <aug>
    <au><snm>Hirsch</snm><fnm>JE</fnm></au>
  </aug>
  <source>Proceedings of the National academy of Sciences</source>
  <publisher>National Acad Sciences</publisher>
  <pubdate>2005</pubdate>
  <volume>102</volume>
  <issue>46</issue>
  <fpage>16569</fpage>
  <lpage>-16572</lpage>
</bibl>

<bibl id="B2">
  <title><p>An improvement of the h-index: The g-index</p></title>
  <aug>
    <au><snm>Egghe</snm><fnm>L</fnm></au>
    <au><cnm>others</cnm></au>
  </aug>
  <source>ISSI newsletter</source>
  <pubdate>2006</pubdate>
  <volume>2</volume>
  <issue>1</issue>
  <fpage>8</fpage>
  <lpage>-9</lpage>
</bibl>

<bibl id="B3">
  <title><p>Universality of scholarly impact metrics</p></title>
  <aug>
    <au><snm>Kaur</snm><fnm>J</fnm></au>
    <au><snm>Radicchi</snm><fnm>F</fnm></au>
    <au><snm>Menczer</snm><fnm>F</fnm></au>
  </aug>
  <source>Journal of Informetrics</source>
  <publisher>Elsevier</publisher>
  <pubdate>2013</pubdate>
  <volume>7</volume>
  <issue>4</issue>
  <fpage>924</fpage>
  <lpage>-932</lpage>
</bibl>

<bibl id="B4">
  <title><p>Finding rising stars in social networks</p></title>
  <aug>
    <au><snm>Daud</snm><fnm>A</fnm></au>
    <au><snm>Abbasi</snm><fnm>R</fnm></au>
    <au><snm>Muhammad</snm><fnm>F</fnm></au>
  </aug>
  <source>International conference on database systems for advanced
  applications</source>
  <pubdate>2013</pubdate>
  <fpage>13</fpage>
  <lpage>-24</lpage>
</bibl>

<bibl id="B5">
  <title><p>Predicting scientific impact based on h-index</p></title>
  <aug>
    <au><snm>Ayaz</snm><fnm>S</fnm></au>
    <au><snm>Masood</snm><fnm>N</fnm></au>
    <au><snm>Islam</snm><fnm>MA</fnm></au>
  </aug>
  <source>Scientometrics</source>
  <publisher>Springer</publisher>
  <pubdate>2018</pubdate>
  <volume>114</volume>
  <issue>3</issue>
  <fpage>993</fpage>
  <lpage>-1010</lpage>
</bibl>

<bibl id="B6">
  <title><p>Learning to predict citation-based impact measures</p></title>
  <aug>
    <au><snm>Weihs</snm><fnm>L</fnm></au>
    <au><snm>Etzioni</snm><fnm>O</fnm></au>
  </aug>
  <source>2017 ACM/IEEE joint conference on digital libraries (JCDL)</source>
  <pubdate>2017</pubdate>
  <fpage>1</fpage>
  <lpage>-10</lpage>
</bibl>

<bibl id="B7">
  <title><p>Predicting long-term scientific impact based on multi-field feature
  extraction</p></title>
  <aug>
    <au><snm>Wu</snm><fnm>Z</fnm></au>
    <au><snm>Lin</snm><fnm>W</fnm></au>
    <au><snm>Liu</snm><fnm>P</fnm></au>
    <au><snm>Chen</snm><fnm>J</fnm></au>
    <au><snm>Mao</snm><fnm>L</fnm></au>
  </aug>
  <source>IEEE Access</source>
  <publisher>IEEE</publisher>
  <pubdate>2019</pubdate>
  <volume>7</volume>
  <fpage>51759</fpage>
  <lpage>-51770</lpage>
</bibl>

<bibl id="B8">
  <title><p>Predicting the citations of scholarly paper</p></title>
  <aug>
    <au><snm>Bai</snm><fnm>X</fnm></au>
    <au><snm>Zhang</snm><fnm>F</fnm></au>
    <au><snm>Lee</snm><fnm>I</fnm></au>
  </aug>
  <source>Journal of Informetrics</source>
  <publisher>Elsevier</publisher>
  <pubdate>2019</pubdate>
  <volume>13</volume>
  <issue>1</issue>
  <fpage>407</fpage>
  <lpage>-418</lpage>
</bibl>

<bibl id="B9">
  <title><p>Predicting citation counts based on deep neural network learning
  techniques</p></title>
  <aug>
    <au><snm>Abrishami</snm><fnm>A</fnm></au>
    <au><snm>Aliakbary</snm><fnm>S</fnm></au>
  </aug>
  <source>Journal of Informetrics</source>
  <publisher>Elsevier</publisher>
  <pubdate>2019</pubdate>
  <volume>13</volume>
  <issue>2</issue>
  <fpage>485</fpage>
  <lpage>-499</lpage>
</bibl>

<bibl id="B10">
  <title><p>HINTS: Citation Time Series Prediction for New Publications via
  Dynamic Heterogeneous Information Network Embedding</p></title>
  <aug>
    <au><snm>Jiang</snm><fnm>S</fnm></au>
    <au><snm>Koch</snm><fnm>B</fnm></au>
    <au><snm>Sun</snm><fnm>Y</fnm></au>
  </aug>
  <source>Proceedings of the Web Conference 2021</source>
  <pubdate>2021</pubdate>
  <fpage>3158</fpage>
  <lpage>-3167</lpage>
</bibl>

<bibl id="B11">
  <title><p>Predicting the citation counts of individual papers via a BP neural
  network</p></title>
  <aug>
    <au><snm>Ruan</snm><fnm>X</fnm></au>
    <au><snm>Zhu</snm><fnm>Y</fnm></au>
    <au><snm>Li</snm><fnm>J</fnm></au>
    <au><snm>Cheng</snm><fnm>Y</fnm></au>
  </aug>
  <source>Journal of Informetrics</source>
  <publisher>Elsevier</publisher>
  <pubdate>2020</pubdate>
  <volume>14</volume>
  <issue>3</issue>
  <fpage>101039</fpage>
</bibl>

<bibl id="B12">
  <title><p>Predicting future citation counts of scientific manuscripts
  submitted for publication: a cohort study in transplantology</p></title>
  <aug>
    <au><snm>Kossmeier</snm><fnm>M</fnm></au>
    <au><snm>Heinze</snm><fnm>G</fnm></au>
  </aug>
  <source>Transplant International</source>
  <publisher>Wiley Online Library</publisher>
  <pubdate>2019</pubdate>
  <volume>32</volume>
  <issue>1</issue>
  <fpage>6</fpage>
  <lpage>-15</lpage>
</bibl>

<bibl id="B13">
  <title><p>Can Author Collaboration Reveal Impact? The Case of
  h-index</p></title>
  <aug>
    <au><snm>Nikolentzos</snm><fnm>G</fnm></au>
    <au><snm>Panagopoulos</snm><fnm>G</fnm></au>
    <au><snm>Evdaimon</snm><fnm>I</fnm></au>
    <au><snm>Vazirgiannis</snm><fnm>M</fnm></au>
  </aug>
  <source>Predicting the Dynamics of Research Impact</source>
  <publisher>Springer</publisher>
  <pubdate>2021</pubdate>
  <fpage>177</fpage>
  <lpage>-194</lpage>
</bibl>

<bibl id="B14">
  <title><p>Academic rising star prediction via scholar’s evaluation model
  and machine learning techniques</p></title>
  <aug>
    <au><snm>Nie</snm><fnm>Y</fnm></au>
    <au><snm>Zhu</snm><fnm>Y</fnm></au>
    <au><snm>Lin</snm><fnm>Q</fnm></au>
    <au><snm>Zhang</snm><fnm>S</fnm></au>
    <au><snm>Shi</snm><fnm>P</fnm></au>
    <au><snm>Niu</snm><fnm>Z</fnm></au>
  </aug>
  <source>Scientometrics</source>
  <publisher>Springer</publisher>
  <pubdate>2019</pubdate>
  <volume>120</volume>
  <issue>2</issue>
  <fpage>461</fpage>
  <lpage>-476</lpage>
</bibl>

<bibl id="B15">
  <title><p>Predicting author h-index using characteristics of the co-author
  network</p></title>
  <aug>
    <au><snm>McCarty</snm><fnm>C</fnm></au>
    <au><snm>Jawitz</snm><fnm>JW</fnm></au>
    <au><snm>Hopkins</snm><fnm>A</fnm></au>
    <au><snm>Goldman</snm><fnm>A</fnm></au>
  </aug>
  <source>Scientometrics</source>
  <publisher>Springer</publisher>
  <pubdate>2013</pubdate>
  <volume>96</volume>
  <issue>2</issue>
  <fpage>467</fpage>
  <lpage>-483</lpage>
</bibl>

<bibl id="B16">
  <title><p>Can scientific impact be predicted?</p></title>
  <aug>
    <au><snm>Dong</snm><fnm>Y</fnm></au>
    <au><snm>Johnson</snm><fnm>RA</fnm></au>
    <au><snm>Chawla</snm><fnm>NV</fnm></au>
  </aug>
  <source>IEEE Transactions on Big Data</source>
  <publisher>IEEE</publisher>
  <pubdate>2016</pubdate>
  <volume>2</volume>
  <issue>1</issue>
  <fpage>18</fpage>
  <lpage>-30</lpage>
</bibl>

<bibl id="B17">
  <title><p>The many facets of academic mobility and its impact on scholars'
  career</p></title>
  <aug>
    <au><snm>Momeni</snm><fnm>F</fnm></au>
    <au><snm>Karimi</snm><fnm>F</fnm></au>
    <au><snm>Mayr</snm><fnm>P</fnm></au>
    <au><snm>Peters</snm><fnm>I</fnm></au>
    <au><snm>Dietze</snm><fnm>S</fnm></au>
  </aug>
  <source>Journal of Informetrics</source>
  <publisher>Elsevier</publisher>
  <pubdate>2022</pubdate>
  <volume>16</volume>
  <issue>2</issue>
  <fpage>101280</fpage>
</bibl>

<bibl id="B18">
  <title><p>Comparing research productivity of returnee-PhDs in science,
  engineering, and the social sciences</p></title>
  <aug>
    <au><snm>Singh</snm><fnm>V</fnm></au>
  </aug>
  <source>Scientometrics</source>
  <publisher>Springer</publisher>
  <pubdate>2018</pubdate>
  <volume>115</volume>
  <issue>3</issue>
  <fpage>1241</fpage>
  <lpage>-1252</lpage>
</bibl>

<bibl id="B19">
  <title><p>What effects does international mobility have on scientists’
  careers? A systematic review</p></title>
  <aug>
    <au><snm>Netz</snm><fnm>N</fnm></au>
    <au><snm>Hampel</snm><fnm>S</fnm></au>
    <au><snm>Aman</snm><fnm>V</fnm></au>
  </aug>
  <source>Research evaluation</source>
  <publisher>Oxford University Press</publisher>
  <pubdate>2020</pubdate>
  <volume>29</volume>
  <issue>3</issue>
  <fpage>327</fpage>
  <lpage>-351</lpage>
</bibl>

<bibl id="B20">
  <title><p>What academic mobility configurations contribute to high
  performance: an fsQCA analysis of CSC-funded visiting scholars</p></title>
  <aug>
    <au><snm>Liu</snm><fnm>J</fnm></au>
    <au><snm>Wang</snm><fnm>R</fnm></au>
    <au><snm>Xu</snm><fnm>S</fnm></au>
  </aug>
  <source>Scientometrics</source>
  <publisher>Springer</publisher>
  <pubdate>2021</pubdate>
  <volume>126</volume>
  <issue>2</issue>
  <fpage>1079</fpage>
  <lpage>-1100</lpage>
</bibl>

<bibl id="B21">
  <title><p>h-Index and Academic Rank by Gender Among Breast Surgery Fellowship
  Faculty</p></title>
  <aug>
    <au><snm>Radford</snm><fnm>DM</fnm></au>
    <au><snm>Parangi</snm><fnm>S</fnm></au>
    <au><snm>Tu</snm><fnm>C</fnm></au>
    <au><snm>Silver</snm><fnm>JK</fnm></au>
  </aug>
  <source>Journal of Women's Health</source>
  <publisher>Mary Ann Liebert, Inc., publishers 140 Huguenot Street, 3rd Floor
  New~…</publisher>
  <pubdate>2022</pubdate>
  <volume>31</volume>
  <issue>1</issue>
  <fpage>110</fpage>
  <lpage>-116</lpage>
</bibl>

<bibl id="B22">
  <title><p>Gender comparisons of social work faculty using h-index
  scores</p></title>
  <aug>
    <au><snm>Carter</snm><fnm>TE</fnm></au>
    <au><snm>Smith</snm><fnm>TE</fnm></au>
    <au><snm>Osteen</snm><fnm>PJ</fnm></au>
  </aug>
  <source>Scientometrics</source>
  <publisher>Springer</publisher>
  <pubdate>2017</pubdate>
  <volume>111</volume>
  <issue>3</issue>
  <fpage>1547</fpage>
  <lpage>-1557</lpage>
</bibl>

<bibl id="B23">
  <title><p>Gender differences in promotion and scholarly impact: an analysis
  of 1460 academic ophthalmologists</p></title>
  <aug>
    <au><snm>Lopez</snm><fnm>SA</fnm></au>
    <au><snm>Svider</snm><fnm>PF</fnm></au>
    <au><snm>Misra</snm><fnm>P</fnm></au>
    <au><snm>Bhagat</snm><fnm>N</fnm></au>
    <au><snm>Langer</snm><fnm>PD</fnm></au>
    <au><snm>Eloy</snm><fnm>JA</fnm></au>
  </aug>
  <source>Journal of surgical education</source>
  <publisher>Elsevier</publisher>
  <pubdate>2014</pubdate>
  <volume>71</volume>
  <issue>6</issue>
  <fpage>851</fpage>
  <lpage>-859</lpage>
</bibl>

<bibl id="B24">
  <title><p>The h index and career assessment by numbers</p></title>
  <aug>
    <au><snm>Kelly</snm><fnm>CD</fnm></au>
    <au><snm>Jennions</snm><fnm>MD</fnm></au>
  </aug>
  <source>Trends in Ecology \& Evolution</source>
  <publisher>Elsevier</publisher>
  <pubdate>2006</pubdate>
  <volume>21</volume>
  <issue>4</issue>
  <fpage>167</fpage>
  <lpage>-170</lpage>
</bibl>

<bibl id="B25">
  <title><p>The relative influences of government funding and international
  collaboration on citation impact</p></title>
  <aug>
    <au><snm>Leydesdorff</snm><fnm>L</fnm></au>
    <au><snm>Bornmann</snm><fnm>L</fnm></au>
    <au><snm>Wagner</snm><fnm>CS</fnm></au>
  </aug>
  <source>Journal of the Association for Information Science and
  Technology</source>
  <publisher>Wiley Online Library</publisher>
  <pubdate>2019</pubdate>
  <volume>70</volume>
  <issue>2</issue>
  <fpage>198</fpage>
  <lpage>-201</lpage>
</bibl>

<bibl id="B26">
  <title><p>A comprehensive analysis of acknowledgement texts in Web of
  Science: a case study on four scientific domains</p></title>
  <aug>
    <au><snm>Smirnova</snm><fnm>N</fnm></au>
    <au><snm>Mayr</snm><fnm>P</fnm></au>
  </aug>
  <source>Scientometrics</source>
  <publisher>Springer</publisher>
  <pubdate>2023</pubdate>
  <volume>128</volume>
  <issue>1</issue>
  <fpage>709</fpage>
  <lpage>-734</lpage>
</bibl>

<bibl id="B27">
  <title><p>Economic, linguistic, and political factors in the scientific
  productivity of countries</p></title>
  <aug>
    <au><snm>Gantman</snm><fnm>ER</fnm></au>
  </aug>
  <source>Scientometrics</source>
  <publisher>Akad{\'e}miai Kiad{\'o}, co-published with Springer Science+
  Business Media BV~…</publisher>
  <pubdate>2012</pubdate>
  <volume>93</volume>
  <issue>3</issue>
  <fpage>967</fpage>
  <lpage>-985</lpage>
</bibl>

<bibl id="B28">
  <title><p>Determinants of citation impact: A comparative analysis of the
  Global South versus the Global North</p></title>
  <aug>
    <au><snm>Confraria</snm><fnm>H</fnm></au>
    <au><snm>Godinho</snm><fnm>MM</fnm></au>
    <au><snm>Wang</snm><fnm>L</fnm></au>
  </aug>
  <source>Research Policy</source>
  <publisher>Elsevier</publisher>
  <pubdate>2017</pubdate>
  <volume>46</volume>
  <issue>1</issue>
  <fpage>265</fpage>
  <lpage>-279</lpage>
</bibl>

<bibl id="B29">
  <title><p>Comparison of the h-index for different fields of research using
  bootstrap methodology</p></title>
  <aug>
    <au><snm>Malesios</snm><fnm>CC</fnm></au>
    <au><snm>Psarakis</snm><fnm>S</fnm></au>
  </aug>
  <source>Quality \& Quantity</source>
  <publisher>Springer</publisher>
  <pubdate>2014</pubdate>
  <volume>48</volume>
  <issue>1</issue>
  <fpage>521</fpage>
  <lpage>-545</lpage>
</bibl>

<bibl id="B30">
  <title><p>The discipline dependence of citation statistics</p></title>
  <aug>
    <au><snm>Lillquist</snm><fnm>E</fnm></au>
    <au><snm>Green</snm><fnm>S</fnm></au>
  </aug>
  <source>Scientometrics</source>
  <publisher>Akad{\'e}miai Kiad{\'o}, co-published with Springer Science+
  Business Media BV~…</publisher>
  <pubdate>2010</pubdate>
  <volume>84</volume>
  <issue>3</issue>
  <fpage>749</fpage>
  <lpage>-762</lpage>
</bibl>

<bibl id="B31">
  <title><p>Scaling the h-index for different scientific ISI fields</p></title>
  <aug>
    <au><snm>Iglesias</snm><fnm>J</fnm></au>
    <au><snm>Pecharrom{\'a}n</snm><fnm>C</fnm></au>
  </aug>
  <source>Scientometrics</source>
  <publisher>Akad{\'e}miai Kiad{\'o}, co-published with Springer Science+
  Business Media BV~…</publisher>
  <pubdate>2007</pubdate>
  <volume>73</volume>
  <issue>3</issue>
  <fpage>303</fpage>
  <lpage>-320</lpage>
</bibl>

<bibl id="B32">
  <title><p>Inequality and cumulative advantage in science careers: a case
  study of high-impact journals</p></title>
  <aug>
    <au><snm>Petersen</snm><fnm>AM</fnm></au>
    <au><snm>Penner</snm><fnm>O</fnm></au>
  </aug>
  <source>EPJ Data Science</source>
  <publisher>Springer</publisher>
  <pubdate>2014</pubdate>
  <volume>3</volume>
  <fpage>1</fpage>
  <lpage>-25</lpage>
</bibl>

<bibl id="B33">
  <title><p>Do open-access dermatology articles have higher citation counts
  than those with subscription-based access?</p></title>
  <aug>
    <au><snm>Xie</snm><fnm>F</fnm></au>
    <au><snm>Ghozy</snm><fnm>S</fnm></au>
    <au><snm>Kallmes</snm><fnm>DF</fnm></au>
    <au><snm>Lehman</snm><fnm>JS</fnm></au>
  </aug>
  <source>PloS one</source>
  <publisher>Public Library of Science San Francisco, CA USA</publisher>
  <pubdate>2022</pubdate>
  <volume>17</volume>
  <issue>12</issue>
  <fpage>e0279265</fpage>
</bibl>

<bibl id="B34">
  <title><p>The Open Access Policy Citation Advantage for a Medical
  School</p></title>
  <aug>
    <au><snm>Blair</snm><fnm>LD</fnm></au>
    <au><snm>Odell</snm><fnm>JD</fnm></au>
  </aug>
  <pubdate>2020</pubdate>
</bibl>

<bibl id="B35">
  <title><p>The post-embargo open access citation advantage: it exists
  (probably), it’s modest (usually), and the rich get richer (of
  course)</p></title>
  <aug>
    <au><snm>Ottaviani</snm><fnm>J</fnm></au>
  </aug>
  <source>PloS one</source>
  <publisher>Public Library of Science San Francisco, CA USA</publisher>
  <pubdate>2016</pubdate>
  <volume>11</volume>
  <issue>8</issue>
  <fpage>e0159614</fpage>
</bibl>

<bibl id="B36">
  <title><p>Investigating the citation advantage of author-pays charges model
  in computer science research: a case study of Elsevier and
  Springer</p></title>
  <aug>
    <au><snm>Amjad</snm><fnm>T</fnm></au>
    <au><snm>Sabir</snm><fnm>M</fnm></au>
    <au><snm>Shamim</snm><fnm>A</fnm></au>
    <au><snm>Amjad</snm><fnm>M</fnm></au>
    <au><snm>Daud</snm><fnm>A</fnm></au>
  </aug>
  <source>Library Hi Tech</source>
  <publisher>Emerald Publishing Limited</publisher>
  <pubdate>2022</pubdate>
  <volume>40</volume>
  <issue>3</issue>
  <fpage>685</fpage>
  <lpage>-703</lpage>
</bibl>

<bibl id="B37">
  <title><p>Is the open access citation advantage real? A systematic review of
  the citation of open access and subscription-based articles</p></title>
  <aug>
    <au><snm>Langham Putrow</snm><fnm>A</fnm></au>
    <au><snm>Bakker</snm><fnm>C</fnm></au>
    <au><snm>Riegelman</snm><fnm>A</fnm></au>
  </aug>
  <source>PloS one</source>
  <publisher>Public Library of Science San Francisco, CA USA</publisher>
  <pubdate>2021</pubdate>
  <volume>16</volume>
  <issue>6</issue>
  <fpage>e0253129</fpage>
</bibl>

<bibl id="B38">
  <title><p>The relationship between bioRxiv preprints, citations and
  altmetrics</p></title>
  <aug>
    <au><snm>Fraser</snm><fnm>N</fnm></au>
    <au><snm>Momeni</snm><fnm>F</fnm></au>
    <au><snm>Mayr</snm><fnm>P</fnm></au>
    <au><snm>Peters</snm><fnm>I</fnm></au>
  </aug>
  <source>Quantitative Science Studies</source>
  <publisher>MIT Press One Rogers Street, Cambridge, MA 02142-1209, USA
  journals-info~…</publisher>
  <pubdate>2020</pubdate>
  <volume>1</volume>
  <issue>2</issue>
  <fpage>618</fpage>
  <lpage>-638</lpage>
</bibl>

<bibl id="B39">
  <title><p>Which Factors Drive Open Access Publishing? A Springer Nature Case
  Study</p></title>
  <aug>
    <au><snm>Momeni</snm><fnm>F</fnm></au>
    <au><snm>Dietze</snm><fnm>S</fnm></au>
    <au><snm>Mayr</snm><fnm>P</fnm></au>
    <au><snm>Biesenbender</snm><fnm>K</fnm></au>
    <au><snm>Peters</snm><fnm>I</fnm></au>
  </aug>
  <source>arXiv preprint arXiv:2208.08221</source>
  <pubdate>2022</pubdate>
</bibl>

<bibl id="B40">
  <title><p>Correlation between impact and collaboration</p></title>
  <aug>
    <au><snm>Hsu</snm><fnm>Jw</fnm></au>
    <au><snm>Huang</snm><fnm>Dw</fnm></au>
  </aug>
  <source>Scientometrics</source>
  <publisher>Akad{\'e}miai Kiad{\'o}, co-published with Springer Science+
  Business Media BV~…</publisher>
  <pubdate>2011</pubdate>
  <volume>86</volume>
  <issue>2</issue>
  <fpage>317</fpage>
  <lpage>-324</lpage>
</bibl>

<bibl id="B41">
  <title><p>International and domestic co-publishing and their citation impact
  in different disciplines</p></title>
  <aug>
    <au><snm>Puuska</snm><fnm>HM</fnm></au>
    <au><snm>Muhonen</snm><fnm>R</fnm></au>
    <au><snm>Leino</snm><fnm>Y</fnm></au>
  </aug>
  <source>Scientometrics</source>
  <publisher>Springer</publisher>
  <pubdate>2014</pubdate>
  <volume>98</volume>
  <issue>2</issue>
  <fpage>823</fpage>
  <lpage>-839</lpage>
</bibl>

<bibl id="B42">
  <title><p>Predicting scientific success based on coauthorship
  networks</p></title>
  <aug>
    <au><snm>Sarig{\"o}l</snm><fnm>E</fnm></au>
    <au><snm>Pfitzner</snm><fnm>R</fnm></au>
    <au><snm>Scholtes</snm><fnm>I</fnm></au>
    <au><snm>Garas</snm><fnm>A</fnm></au>
    <au><snm>Schweitzer</snm><fnm>F</fnm></au>
  </aug>
  <source>EPJ Data Science</source>
  <publisher>Springer</publisher>
  <pubdate>2014</pubdate>
  <volume>3</volume>
  <fpage>1</fpage>
  <lpage>-16</lpage>
</bibl>

<bibl id="B43">
  <title><p>Relationship between international collaboration papers and their
  citations from an economic perspective</p></title>
  <aug>
    <au><snm>Ni</snm><fnm>P</fnm></au>
    <au><snm>An</snm><fnm>X</fnm></au>
  </aug>
  <source>Scientometrics</source>
  <publisher>Springer</publisher>
  <pubdate>2018</pubdate>
  <volume>116</volume>
  <issue>2</issue>
  <fpage>863</fpage>
  <lpage>-877</lpage>
</bibl>

<bibl id="B44">
  <title><p>Inferring gender from names on the web: A comparative evaluation of
  gender detection methods</p></title>
  <aug>
    <au><snm>Karimi</snm><fnm>F</fnm></au>
    <au><snm>Wagner</snm><fnm>C</fnm></au>
    <au><snm>Lemmerich</snm><fnm>F</fnm></au>
    <au><snm>Jadidi</snm><fnm>M</fnm></au>
    <au><snm>Strohmaier</snm><fnm>M</fnm></au>
  </aug>
  <source>Proceedings of the 25th International conference companion on World
  Wide Web</source>
  <pubdate>2016</pubdate>
  <fpage>53</fpage>
  <lpage>-54</lpage>
</bibl>

<bibl id="B45">
  <title><p>From P100 to P100': A new citation-rank approach</p></title>
  <aug>
    <au><snm>Bornmann</snm><fnm>L</fnm></au>
    <au><snm>Mutz</snm><fnm>R</fnm></au>
  </aug>
  <source>Journal of the Association for Information Science and
  Technology</source>
  <publisher>Wiley Online Library</publisher>
  <pubdate>2014</pubdate>
  <volume>65</volume>
  <issue>9</issue>
  <fpage>1939</fpage>
  <lpage>-1943</lpage>
</bibl>

<bibl id="B46">
  <title><p>An evaluation of percentile measures of citation impact, and a
  proposal for making them better</p></title>
  <aug>
    <au><snm>Bornmann</snm><fnm>L</fnm></au>
    <au><snm>Williams</snm><fnm>R</fnm></au>
  </aug>
  <source>Scientometrics</source>
  <publisher>Springer</publisher>
  <pubdate>2020</pubdate>
  <volume>124</volume>
  <issue>2</issue>
  <fpage>1457</fpage>
  <lpage>-1478</lpage>
</bibl>

<bibl id="B47">
  <title><p>Xgboost: A scalable tree boosting system</p></title>
  <aug>
    <au><snm>Chen</snm><fnm>T</fnm></au>
    <au><snm>Guestrin</snm><fnm>C</fnm></au>
  </aug>
  <source>Proceedings of the 22nd acm sigkdd international conference on
  knowledge discovery and data mining</source>
  <pubdate>2016</pubdate>
  <fpage>785</fpage>
  <lpage>-794</lpage>
</bibl>

<bibl id="B48">
  <title><p>Using the R-MAPE index as a resistant measure of forecast
  accuracy</p></title>
  <aug>
    <au><snm>Blasco</snm><fnm>BC</fnm></au>
    <au><snm>Moreno</snm><fnm>JJM</fnm></au>
    <au><snm>Pol</snm><fnm>AP</fnm></au>
    <au><snm>Abad</snm><fnm>AS</fnm></au>
  </aug>
  <source>Psicothema</source>
  <publisher>Colegio Oficial De Psicologos Del Principado De
  Asturias</publisher>
  <pubdate>2013</pubdate>
  <volume>25</volume>
  <issue>4</issue>
  <fpage>500</fpage>
  <lpage>-506</lpage>
</bibl>

<bibl id="B49">
  <title><p>Will this paper increase your h-index? Scientific impact
  prediction</p></title>
  <aug>
    <au><snm>Dong</snm><fnm>Y</fnm></au>
    <au><snm>Johnson</snm><fnm>RA</fnm></au>
    <au><snm>Chawla</snm><fnm>NV</fnm></au>
  </aug>
  <source>Proceedings of the eighth ACM international conference on web search
  and data mining</source>
  <pubdate>2015</pubdate>
  <fpage>149</fpage>
  <lpage>-158</lpage>
</bibl>

<bibl id="B50">
  <title><p>Review the performance of the Bernoulli Na{\"\i}ve Bayes Classifier
  in Intrusion Detection Systems using Recursive Feature Elimination with
  Cross-validated selection of the best number of features</p></title>
  <aug>
    <au><snm>Artur</snm><fnm>M</fnm></au>
  </aug>
  <source>Procedia Computer Science</source>
  <publisher>Elsevier</publisher>
  <pubdate>2021</pubdate>
  <volume>190</volume>
  <fpage>564</fpage>
  <lpage>-570</lpage>
</bibl>

<bibl id="B51">
  <title><p>RFE Based Feature Selection Improves Performance of Classifying
  Multiple-causes Deaths in Colorectal Cancer</p></title>
  <aug>
    <au><snm>Zhao</snm><fnm>L</fnm></au>
    <au><snm>Deng</snm><fnm>F</fnm></au>
    <au><snm>Zhang</snm><fnm>X</fnm></au>
    <au><snm>Yu</snm><fnm>N</fnm></au>
  </aug>
  <source>2022 7th International Conference on Intelligent Informatics and
  Biomedical Science (ICIIBMS)</source>
  <pubdate>2022</pubdate>
  <volume>7</volume>
  <fpage>188</fpage>
  <lpage>-194</lpage>
</bibl>

<bibl id="B52">
  <title><p>Statistics for Business and Economics</p></title>
  <aug>
    <au><snm>Newbold</snm><fnm>P</fnm></au>
    <au><snm>Carlson</snm><fnm>WL</fnm></au>
    <au><snm>Thorne</snm><fnm>B</fnm></au>
  </aug>
  <publisher>Pearson</publisher>
  <pubdate>2013</pubdate>
</bibl>

<bibl id="B53">
  <title><p>Increasing importance of research metrics: Journal Impact Factor
  and h-index</p></title>
  <aug>
    <au><snm>Grech</snm><fnm>V</fnm></au>
    <au><snm>Rizk</snm><fnm>DE</fnm></au>
  </aug>
  <source>International Urogynecology Journal</source>
  <publisher>Springer</publisher>
  <pubdate>2018</pubdate>
  <volume>29</volume>
  <fpage>619</fpage>
  <lpage>-620</lpage>
</bibl>

<bibl id="B54">
  <title><p>Theory and practise of the g-index</p></title>
  <aug>
    <au><snm>Egghe</snm><fnm>L</fnm></au>
  </aug>
  <source>Scientometrics</source>
  <publisher>Citeseer</publisher>
  <pubdate>2006</pubdate>
  <volume>69</volume>
  <issue>1</issue>
  <fpage>131</fpage>
  <lpage>-152</lpage>
</bibl>

<bibl id="B55">
  <title><p>Detecting h-index manipulation through self-citation
  analysis</p></title>
  <aug>
    <au><snm>Bartneck</snm><fnm>C</fnm></au>
    <au><snm>Kokkelmans</snm><fnm>S</fnm></au>
  </aug>
  <source>Scientometrics</source>
  <publisher>Akad{\'e}miai Kiad{\'o}, co-published with Springer Science+
  Business Media BV~…</publisher>
  <pubdate>2011</pubdate>
  <volume>87</volume>
  <issue>1</issue>
  <fpage>85</fpage>
  <lpage>-98</lpage>
</bibl>

<bibl id="B56">
  <title><p>AI ethics in predictive policing: From models of threat to an
  ethics of care</p></title>
  <aug>
    <au><snm>Asaro</snm><fnm>PM</fnm></au>
  </aug>
  <source>IEEE Technology and Society Magazine</source>
  <publisher>IEEE</publisher>
  <pubdate>2019</pubdate>
  <volume>38</volume>
  <issue>2</issue>
  <fpage>40</fpage>
  <lpage>-53</lpage>
</bibl>

<bibl id="B57">
  <title><p>Discrimination, artificial intelligence, and algorithmic
  decision-making</p></title>
  <aug>
    <au><snm>Zuiderveen Borgesius</snm><fnm>F</fnm></au>
    <au><cnm>others</cnm></au>
  </aug>
  <source>l{\'\i}nea], Council of Europe</source>
  <pubdate>2018</pubdate>
</bibl>

</refgrp>
} 




\section*{Figure Legends}
\autoref{fig:heatmap}: \nameref{fig:heatmap}

\vspace{5pt}
\autoref{fig:performance_methods}: \nameref{fig:performance_methods}

\vspace{5pt}
\autoref{figure_mape_seniorit}: \nameref{figure_mape_seniorit}

\vspace{5pt}

\section*{Table Legends}
\autoref{TableStatisticAuthorPaper}: \nameref{TableStatisticAuthorPaper}

\vspace{5pt}
\autoref{TableFeatures}: \nameref{TableFeatures}

\vspace{5pt}
\autoref{TableFeatureDescription}: \nameref{TableFeatureDescription}

\vspace{5pt}
\autoref{TableFeatCombi}: \nameref{TableFeatCombi}

\vspace{5pt}
\autoref{table_corr}: \nameref{table_corr}

\vspace{5pt}
\autoref{tableFeatureSelectRFE}: \nameref{tableFeatureSelectRFE}

\vspace{5pt}
\autoref{TablePerformanceHindex}: \nameref{TablePerformanceHindex}

\vspace{5pt}

\end{backmatter}
\end{document}